\RequirePackage{fix-cm}
\documentclass[twocolumn]{svjour3}          
\smartqed  
\usepackage{graphicx}
\usepackage{bm}
\usepackage{amsmath}
\usepackage{amssymb}
\usepackage{mathptmx}      
\usepackage{latexsym}
\usepackage{verbatim}
\usepackage[ansinew]{inputenc}
\usepackage{color}
\usepackage[english]{babel}
\usepackage{hyperref}
\newcommand{\be}    {\begin{equation}}
\newcommand{\ee}    {\end{equation}}
\newcommand{\ba}   {\begin{eqnarray}}
\newcommand{\ea}   {\end{eqnarray}}

\begin{document}
\title{The recursive Green's function method for graphene}


\author{Caio H. Lewenkopf \and 
  Eduardo R. Mucciolo }


\institute{Caio H. Lewenkopf  \at
           Instituto de F\'{\i}sica, Universidade Federal Fluminense, Brazil \\
           \email{caio@if.uff.br} 
           \and
Eduardo R. Mucciolo  \at
               Department of Physics, University of Central Florida, USA  \\
              \email{mucciolo@physics.ucf.edu}}

\date{Received: date / Accepted: date}

\maketitle

\begin{abstract}
We describe how to apply the recursive Green's function method to the
computation of electronic transport properties of graphene sheets and
nanoribbons in the linear response regime. This method allows for an
amenable inclusion of several disorder mechanisms at the microscopic
level, as well as inhomogeneous gating, finite temperature, and, to
some extend, dephasing. We present algorithms for computing the
conductance, density of states, and current densities for armchair and
zigzag atomic edge alignments. Several numerical results are presented
to illustrate the usefulness of the method.

\end{abstract}

\section{Introduction}
\label{sec:intro}

Since graphene was first produced, several synthesis strategies have
been put forward. Significant progress has been made to produce better
quality samples with the goal of improving their transport
properties. Despite the enormous effort, we are still very far from
reaching the perfect ballistic regime and disorder always plays a
central role, particularly in electronic transport. Disorder appears
in several different forms, being either local (such as lattice
defects, edge irregularities, and surface adsorbates) or long ranged
(such as charge impurities trapped in the substrate or ripples due to
substrate roughness) \cite{CastroNeto09}.

Several theoretical methods have been developed to describe electronic
transport in disordered graphene \cite{Mucciolo10}. The effective
low-energy Dirac Hamiltonian, derived from band-structure theory,
combined with a standard diagrammatic perturbative expansion is an
excellent analytical tool for giving us insight into the properties of
disordered graphene \cite{Shon98,Ostrovsky06}. However, it has $(k_F
\ell)^{-1}$ as a small expansion parameter, where $k_F$ stands for the
Fermi wave number and $\ell$ is the electron mean free path. Thus, it
describes well the conductivity in graphene at high doping, but
becomes of limited use when one is interested in the physics close to
the charge neutrality point, where $k_F \ell \ll 1$. Theoretical
investigations of the that regime require instead the use of numerical
methods.

Most numerical methods employed to study the transport properties of
disordered graphene use an atomistic basis \cite{Mucciolo10}. The few
exceptions are tailor-made methods to deal with long-range disorder,
where either a momentum representation \cite{Nomura07,Nomura07b} or
discretized version of the Dirac equation
\cite{Tworzydlo08,Hernandez12} are used within a single-valley
approximation.

For many applications, one is interested in the two- or multiple-probe
conductance. For the conductance, differently from the conductivity,
geometry plays an important role. The recursive Green's function (RGF)
method \cite{Thouless81} became the standard tool to compute transport
properties in this case. The method is very reliable, computationally
efficient, and allows for a parallel implementation
\cite{Drouvelis06}. It can model arbitrary geometries and efficiently
addresses a variety of scattering processes within the single-particle
approximation. The goal of this paper is to show how to compute
electronic transport properties of graphene samples within the
tight-binding approximation using the RGF method. The key element is
an efficient algorithm for evaluating the single-particle Green's
function of sheets or ribbons.

The recursive method was developed by Thouless and Kirkpatrick
\cite{Thouless81} for computing the linear electronic conductance of
linear atomic chains in the presence of on-site disorder. The method
was later generalized to two-dimensional systems in the ``slice"
formulation, which is the form most used nowadays
\cite{MacKinnon85}. Variations of the method have been introduced in
the literature to treat three-dimensional \cite{Sols89} and
multi-probe systems \cite{Baranger91} with arbitrary geometries, see
e.g. Ref. \cite{Kazymyrenko08}.

We note that other efficient, atomistic methods have been employed in
recent years to study electronic transport in mesoscopic systems: For
instance, the wave-packet time evolution \cite{Kramer10,Yuan10}, the
kernel polynomial expansion \cite{Weisse06,Ferreira11}, and the
continued fraction expansion \cite{Triozon05}, to name a
few. Recently, an alternative method to compute transport of ballistic
graphene junctions, particularly effective when strong magnetic fields
are presented, was introduced \cite{Liu2012}. However, for most
practitioners, the RGF remains the best method for tackling
large-scale but finite-size problems where quantum coherence and
disorder are present simultaneously.

This paper does not attempt to be a comprehensive review of the
recursive method, but rather a self-contained description that gives
to interested readers, yet unfamiliar with the RGF method, all the
basic material necessary to implement a calculation on their own. For
that purpose, we briefly present some standard material covered in
textbooks \cite{Datta96,Ferry97}, discuss some more advanced issues
which are found scattered in the literature, and present original
developments tailor-made for graphene.

This paper is organized as follows. We begin by quickly reviewing some
fundamental relations of electronic transport theory and by providing
the essential formulation of the method. In Sec.~\ref{sec:RGFmethod}
we present the recursive Green's function method. The method requires
as input the surface Green's functions of the electronic leads, taken
at the lead-device interface. In Sec.~\ref{sec:contacts} we describe
how to compute the surface Green's function of semi-infinite lattices
that play the role of leads. Next, we present an efficient
discretization scheme to implement the recursive method for graphene
sheets and nanoribbons. In Sec. \ref{sec:localquant} we show how to
evaluate quantities such as the local density of states and local
current densities. Very often, one is interested in cases where the
coherence length $\ell_\phi$ is comparable to the system size $L$. For
such situations, it is possible to account for dephasing using the
phenomenological voltage probe model, as described in
Sec.~\ref{sec:dephasing}. In the context of graphene, the main
application of the RGF method is the study of disorder effects in
electronic transport. We discuss the main kinds of disorder and show
how to account for them in Sec.~\ref{sec:disorder}. We conclude by
presenting a number of numerical results that illustrate the method in
Sec.~\ref{sec:numerics}.

\section{Elements of Linear Mesoscopic Transport}
\label{sec:mesoscopic}

In this Section we review the key elements necessary to implement the
recursive Green's function method for two-dimensional systems. The
linear dc conductance is computed using the exact single-particle
retarded Green's function that connects the source and drain leads, in
conjunction with either the Landauer \cite{Landauer88} or the Caroli
formula \cite{Caroli71}.

For a two-probe setup, as illustrated by Fig.~\ref{fig:layout}, the
zero-temperature linear conductance is given by the Landauer formula
\begin{equation}
{\cal G} = \frac{2e^2}{h} {\rm Tr}_c \left[ t^\dagger t \right].  \ee
Here, $t (t^\prime)$ is the transmission matrix {\it across} the
system from left to right (right to left) and $r (r^\prime)$ is the
reflection matrix at the left-hand (right-hand) side. The factor of 2
stands for spin degeneracy and the trace is taken over the propagating
modes at the left and right leads. The transmission matrix can be
obtained from the $S$ matrix,
\begin{equation} 
S = \left(\begin{array}{cc} r & t^\prime \\ t & r^\prime
\end{array}\right), 
\end{equation}
which is given by \cite{Lee81,Fisher81}
\begin{align}
\label{eq:contSmatrix}
S_{ab}(E) =& - \delta_{ab} + i \hbar \sqrt{v_av_b} \nonumber \\ &
\;\;\;\times \int \! dy_q \int \! dy_p \, \chi_a^*(y_q)G^r(y_q,y_p;E)
\chi_b(y_p),
\end{align}
where $v_c$ and $\chi_c(y_p)$ are, respectively, the longitudinal
propagation velocity and its transverse wave function in the
propagating channel $c$ of lead $p$ (either on the left-hand or
right-hand side). The integrations run over the contact regions at the
right and left terminations of the graphene sheet (see
Fig. \ref{fig:layout}). The key element in Eq. (\ref{eq:contSmatrix})
is $G^r(y_q,y_p;E)$, the retarded Green's function corresponding to an
electron with energy $E$ propagating from positions $y_p$ to $y_q$.

\begin{figure}[h]
\centering
\includegraphics[width=\columnwidth]{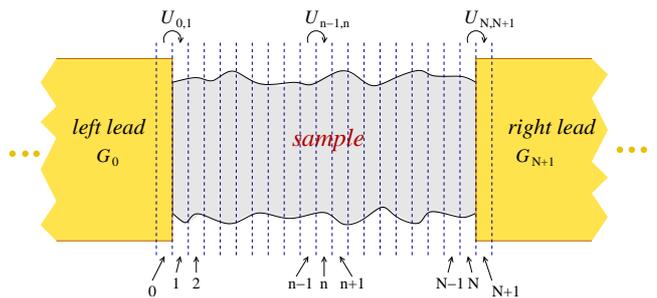}
\vskip0.4cm
\caption{Typical two-probe scheme used in the numerical calculations:
  The sample is described by $G(y_q,y_p;E)$. The perfect leads can be
  accounted for by either propagating mode wave functions $\chi_c$ and
  their density of states $\rho_c$ (Landauer formula) or by level
  widths $\Gamma$ (Caroli formula).}
\label{fig:layout}
\end{figure}

The recursive Green's function method reviewed in this paper is an
efficient tool to compute the scattering properties of noninteracting
electrons described by a tight-binding Hamiltonian of the form
\be 
\label{eq:tight-bindingH}
H = -\sum_{i\neq j} \left( t_{ij}|i\rangle\langle j| + {\rm
  H.c}\right) + \sum_i V_i |i\rangle\langle i|.
\ee
In graphene, the hopping terms $t_{ij}$ typically connect only nearest
neighbor sites $i$ and $j$ of a honeycomb lattice (single-orbital
approximation), although it is straightforward to include
next-to-nearest hopping terms as well, if required. The model can
account for an external magnetic field by a suitable modification of
the hopping terms, as shown in Appendix \ref{sec:A}. Here, $V_i$
stands for a local potential due to gating or disorder.

For simplicity, here we consider only tight-binding models with
orthogonal orbitals and nearest-neighbor hopping terms. The
computational cost of the RGF method scales as $N \times M^3$, where
$N$ is the number of slices and $M$ is the typical number of sites in
a given slice, see Fig. \ref{fig:layout}. The RGF implementation
scheme presented in this paper is particularly recommended when the
system's translational invariance is broken by disorder and/or an
irregular geometry. For systems with translational invariance at the
transverse direction, it can be advantageous to work in the $k$-space
and use alternative hybrid RGF implementations (such as in
Ref. \cite{Sajjad13}) or other methods, e.g. Ref. \cite{Liu2012}.

In the tight-binding basis, Eq.~(\ref{eq:contSmatrix}) reads
\be
\label{eq:tbSmatrix}
S_{ab}(E) = - \delta_{ab} + i \hbar \frac{\sqrt{v_av_b}}{a_0} \sum_{i\in p}
\sum_{j\in q} \chi_a^*(i)G^r_{qp}(i,j) \chi_b(j),
\ee
where the sums run over the sites at the contacts $p$ and $q$ where
the propagating channels $a$ and $b$ are defined, respectively. Notice
that in two spatial dimensions
\begin{align}
& \chi_a(y_p)\rightarrow \frac{1}{\sqrt{a_0}}\chi_a(i)=
\frac{1}{\sqrt{a_0}}\langle i | \chi_a\rangle \\
\mbox{and} \nonumber \\
&G^r(y_q,y_p) \rightarrow \frac{1}{a_0^2}G^r_{qp}(j,i),
\end{align}
where $a_0$ is the lattice constant. Let us consider the case where
the leads are modeled by semi-infinite square lattices. One can then
introduce the level widths \cite{Datta96}
\be
\label{eq:tbGamma}
\Gamma_p(i, i^\prime) = \sum_a \chi_a(i) \frac{\hbar v_a}{a_0}
\chi_a(i^\prime).
\ee
It is straightforward to show that, in this case,
\begin{equation}
\label{eq:transmission}
\mbox{Tr}_c \left[t t^\dagger \right] = \sum_{{i,i^\prime \in
    L}\atop{j,j^\prime \in R}} \Gamma_L(i,i^\prime) G^r_{LR}(i^\prime,
j)\Gamma_R(j,j^\prime)G^a_{RL}(j^\prime,i) = {\cal T} 
\end{equation}
where
\begin{equation}
{\cal T} \equiv \mbox {Tr}_s \Big[\Gamma_L G_{LR}^r \Gamma_R G_{RL}^a
  \Big].
 \end{equation}
Here, the subscript in the trace indicates whether the sums run over
channels ($c$) or sites ($s$).  Depending on the author, the
expression on the r.h.s. of Eq.~(\ref{eq:transmission}) is called
either Caroli \cite{Caroli71} or Meir-Wingreen \cite{Meir92}
conductance formula.

This demonstration of the equivalence between the Landauer and Caroli
formulas relies on the Fisher and Lee $S$-matrix {\sl and} on an
expression for $\Gamma_p$ which is only suitable for a square
lattice. This derivation is simple and to some extend non rigorous
but captures the essential elements that will be discussed in what
follows, namely, the Green's functions $G^{(r,a)}_{RL}$ and the decay
width matrices $\Gamma_{R,L}$. There are several ways to show that
\eqref{eq:transmission} holds in general in the linear response
regime, see e.g. Ref. \cite{Hernandez07}.

When the full $S$ matrix is known, it is possible to obtain the {\it
  global} density of states through the Wigner time delay
\cite{Lewenkopf08}, namely,
\be 
\label{eq:DOS-WS}
\rho(E) = - \frac{i}{2\pi} \mbox{Tr}_c \left(S^\dagger
\frac{\partial S}{\partial E}\right), 
\ee
where the derivative of $S$ with respect to the energy can be done
numerically. The computation of \eqref{eq:DOS-WS} is significantly
less expensive than evaluating $\rho(E)$ through the standard
expression, namely,
\be 
\label{eq:DOS-traceG}
\rho(E) = -\frac{1}{\pi} \mbox{Im} \Big[ \mbox{Tr}_{s'}\,
  G^r(E) \Big],
\ee
but it requires the knowledge of the explicit form of the lead wave
functions $\chi_a(i)$. Note that, in Eq.~\eqref{eq:DOS-traceG}, the
trace is taken over all sites of the graphene sample.

The Fano factor is another quantity of interest \cite{Lewenkopf08}. It
can be evaluated through the expression
\begin{equation}
\label{eq:fano1}
F = 1 - \frac{{\rm Tr}_c \left[ t^\dagger t\, t^\dagger t
    \right]}{{\rm Tr}_c \left[ t^\dagger t \right]}.
\end{equation}
Notice that one can define left-to-right and right-to-left Fano
factors, as in the case for the conductance, by switching the matrix
$t$ with $t^\prime$. The Fano factor can be computed without an
explicit knowledge of the wave functions $\chi_a(i)$ by noticing that
Eq. (\ref{eq:fano1}) can be recast as
\begin{equation}
\label{eq:fano2}
F = 1 - \frac{\mbox {Tr}_s \Big[\Gamma_L G_{LR}^r \Gamma_R G_{RL}^a
    \Gamma_L G_{LR}^r \Gamma_R G_{RL}^a \Big]}{\cal T}.
\end{equation}

Let us present the same basic expression in a more suitable form for
the recursive calculations. We begin by writing the Caroli formula for
the transmission probability at a given energy $E$ using the slice
indexing,
\begin{equation}
\label{eq:caroli}
{\cal T} = {\rm Tr}_s \left[ \Gamma_L\, G^r_{0,N+1}(E)\, \Gamma_R\,
  G^a_{N+1,0}(E) \right],
\end{equation}
where $G^{r,a}$ are the retarded and advanced Green's functions across
the system (see Fig. \ref{fig:slices} for a definition of the
subscripts in terms of slice numbers). These Green's functions are
matrices whose rank is defined by the number of sites in the
slices.\footnote{The number of sites per slice does not need to be
  equal for all slices.} The level width matrices are given by the
expression
\begin{equation}
\label{eq:gammas}
\Gamma_{L,R} = i \left[ \Sigma_{L,R}^r(E) -
  \Sigma_{L,R}^{r\,\dagger}(E) \right],
\end{equation}
where the retarded {\it surface} self-energies of the leads read
\begin{equation}
\label{eq:sigmas}
\Sigma_L^r(E) = u_L\, g_L^r(E)\, u_L^\dagger \quad \mbox{and} \quad
\Sigma_R^r(E) = u_R^\dagger\, g_R^r(E)\, u_R.
\end{equation}
Notice that the retarded Green's functions $g_L^r$ and $g_R^r$ are
defined at the surface of the left and right leads, respectively, when
the leads are decoupled from the system. They obey the self-consistent
equations
\begin{align}
\label{eq:gRL}
& \left[ E + i0^+ - h_L - \Sigma^r_L(E) \right] g_L^r(E) = I 
\quad \mbox{and} \quad \nonumber\\
& \left[ E + i0^+ - h_R - \Sigma^r_R(E) \right] g_R^r(E) = I,
\end{align}
where $h_L$ and $h_R$ are the Hamiltonians of isolated, individual
slices in the left and right leads, respectively. The connection
matrices $u_L$ and $u_R$ are defined to run from left-to-right and are
assumed uniform inside the leads. If the leads are identical, then
$g_R^r = g_L^r$, $h_L = h_R$, $u_R = u_L^\dagger$, $\Sigma^r_R(E) =
\Sigma^r_L(E)$, and $\Gamma_R = \Gamma_L$. Notice that, in general,
the coupling matrices are Hermitian, $\left( \Gamma_{R,L}
\right)^\dagger = \Gamma_{R,L}$, while $\left(G^r\right)^\dagger =
G^a$. These two properties guarantee that the transmission probability
computed with Eq. (\ref{eq:caroli}) is always real. Moreover, since
these coupling matrices are also positive by their definition in
Eq. (\ref{eq:gammas}) [notice that the imaginary part of the retarded
  self-energy is negative if we adopt Eq. (\ref{eq:gRL})], one can
show that the transmission probability is positive, as it should be.

\begin{figure}[ht]
\centering
\includegraphics[width=0.90\columnwidth]{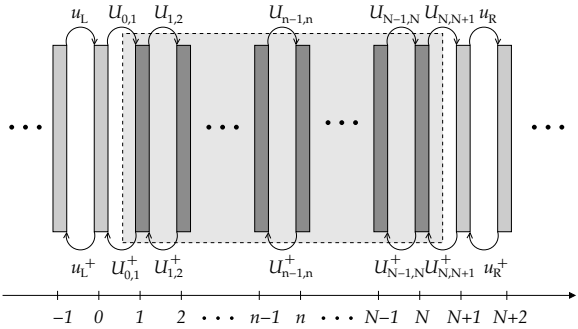}
\caption{Slicing scheme. The central rectangle containing the dark
  strips (slices) represents the bulk of the sample}
\label{fig:slices}
\end{figure}

\section{Recursive Green's Functions}
\label{sec:RGFmethod}

We now present the recursive Green's function method, a very efficient
way to compute the Green's functions that appear in
Eq.~(\ref{eq:caroli}). We begin by introducing two equivalent Dyson
formulas for an exact Green's function (for a derivation of these
formulas, see Refs. \cite{Datta96,HaugJauho08}),
\begin{eqnarray}
\label{eq:dyson}
G & = & G^{(0)} + G^{(0)}\, V\, G, \\
\label{eq:dyson2}
G & = & G^{(0)} + G\, V\, G^{(0)},
\end{eqnarray}
where $G^{(0)}$ represents the ``unperturbed'' Green's function and
$V$ the perturbation. We use these expressions to obtain recursive
relations for the exact Green's function of a quasi-one-dimensional
system coupled to leads. The basic idea is to break up the system into
independent parts (leads and slices) and associate to these parts
``unperturbed'' Green's functions $G^{(0)}$. The hoping matrix
elements connecting those parts are then selectively built into the
perturbation $V$. By choosing the connecting matrix elements and
applying Eqs. (\ref{eq:dyson}) and (\ref{eq:dyson2}) judiciously, we
can build the full Green's function $G$ slice by slice.

Our presentation is specialized to the case of two-probe conductance,
see Fig.~\ref{fig:layout}. It is necessary to derive several
intermediate recurrence formulas before obtaining expressions for the
exact Green's function. We first run the recurrence from left to
right, generating a family of Green's functions $G^L$. Thus, at every
step, Eq. (\ref{eq:dyson}) is employed using a different choice for
$G^{(0)}$ and ${V}$. We repeat the procedure from right to left,
generating another set of functions $G^R$. Finally, we join these two
families to obtain the exact $G$ for the whole system. In this way,
all parts and connecting matrix elements are used (and never double
counted).

The system is broken into $N$ thin slices, each one carrying a maximum
of $M$ sites or cells, as show in Fig. \ref{fig:slices}. The slices
with numbers lower than 1 or larger than $N$ represent the left and
right leads, respectively. The corresponding retarded surface Green's
functions (when the leads are decoupled from the system) are denoted
by $g_L(E)$ and $g_R(E)$, as noted earlier. These Green's functions
are computed separately and before the recurrence procedure (see
Sec.~\ref{sec:contacts}). The retarded Green's function of the
isolated $n$th slice in the system, $g_n(E) = \left( E - h_n +
i0^+\right)^{-1}$, does not need to be individually evaluated before
the recursive calculations. Here, $h_n$ denotes the Hamiltonian of the
isolated $n$th slice.

Neighboring slices within the sample are connected to each other
through the matrices $U_{n-1,n}$ (left to right) and $[U_{n-1,n}]^\dagger
\equiv U_{n,n-1}$ (right to left), with $n=1,\ldots,N$. The first and
last slices in the system are connected to their nearest neighboring
slices in the leads through the coupling matrices $U_{0,1}$ and
$U_{N,N+1}$. The matrix elements of these matrices are the
tight-binding hopping amplitudes connecting sites at different slices.

Here, we assumed that the matrices $U$ only connect nea\-rest-neighbor
slices. For tight-binding models that include next-nearest hopping
terms, one can still use this algorithm by doubling the ``width" of
the unit slices, which slows down the computation by a factor
$2^3$. It also is possible to deal with next-nearest hopping terms and
incur in a smaller slowdown factor by properly modifying the recursive
method \cite{MuccioloXX}.

We use subscripts to denote longitudinal spatial indices (except for
$g_L$, $g_R$, and $g_n$). Thus, $G_{n,m}(E)$ is the matrix Green's
function connecting the $n$ and $m$ slices. Sites indices are shown as
a pair of variables: $G_{n,m}(j,j^\prime)$ denotes the Green's
function connecting site $j$ in the $n$th slice to site $j^\prime$ in
the $m$th slice. Hereafter, we will drop the energy variable $E$
(since scattering is assumed elastic, $E$ is conserved throughout the
system).

\subsection{Connection to leads}
\label{sec:leads}

For the two-terminal setup we address here, the sample (central
region) is coupled to a left lead $L$ and to a right lead $R$. In the
following we show how to built the Green's function that describes
this coupling.

We begin finding the Green's function $G^L$. We recall that the
rightmost slice of the left lead if denoted by 0. Our goal is to
obtain $G^L_{0,n}$ and $G^L_{n,n}$ in order to describe electron
propagation in the sample when the left lead is taken into
account. The reason will become clear when we reach Sec. \ref{sec:full}

The first step is to incorporate the $n=1$ slice to the left contact
Green's function $g_L$. This kind of operation is repeated throughout
the method and therefore we present it in detail. For this purpose, we
introduce the kets $|0\rangle$ and $|1\rangle$ which represent the
states where electrons are found in slices $n=0$ and $n=1$,
respectively. The ``unperturbed" Green's function in this case is
$G^{(0)} = |0\rangle g_L \langle 0| + |1\rangle g_1 \langle 1|$ while
$V = |0\rangle U_{0,1}\langle 1| + |1\rangle U_{1,0}\langle 0|$ is the
perturbation that connects the $n=1$ slice to the left lead. Then,
using Eq. (\ref{eq:dyson}), we obtain
\begin{eqnarray}
\langle 1| G^L|1\rangle & = & \langle 1| G^{(0)}|1 \rangle +
\sum_{m,m^\prime} \langle 1| G^{(0)} |m\rangle \langle m |V
|m^\prime\rangle \langle m'| G^L | 1\rangle \nonumber \\ & = & \langle
1| G^{(0)}|1 \rangle + \langle 1| G^{(0)} |1\rangle \langle 1 |V
|0\rangle \langle 0| G^L | 1\rangle
\end{eqnarray}
and
\begin{eqnarray}
\langle 0 | G^L | 1 \rangle & = & \langle 0| G^{(0)}|1 \rangle +
\sum_{m,m'} \langle 0 | G^{0}|m\rangle \langle m| V | m'\rangle
\langle m| G^L | 1\rangle \nonumber \\ & = & \langle 0 |
G^{0}|0\rangle \langle 0| V | 1\rangle \langle 1| G^L | 1\rangle.
\end{eqnarray}
Adopting the more compact notation $\langle n | G^L | m \rangle =
G^L_{n,m}$, we drop the bras and kets and can rewrite these equation
as
\begin{equation}
G_{1,1}^L = g_1 + g_1 U_{1,0} G^L_{0,1}
\end{equation}
and
\begin{equation} 
G_{0,1}^L = g_L\, U_{0,1}\, G^L_{1,1}.
\end{equation}
Therefore,
\begin{equation}
G_{1,1}^L = \left( I - g_1\, U_{1,0}\, g_L\, U_{0,1} \right)^{-1} g_1.
\end{equation}
Now, since $g_1 = \left(E-h_1 \right)^{-1}$, we can write
\begin{equation}
\label{eq:leadL}
 G_{1,1}^L = \left( E - h_1 - U_{1,0}\, g_L\,
U_{0,1} \right)^{-1}.
\end{equation}
Notice that this Green's function takes into account the coupling of
the first slice with the left lead, but has no information about the
rest of the system or the right lead.

It is important to remark that we neglected the infinitesimal
imaginary part in $g_1$ because we assumed that the ``self-energy''
term in Eq. (\ref{eq:leadL}) brings its own {\it finite} imaginary
part.

We proceed analogously in order to connect the last slice to the right
lead. Choosing $G^{(0)} = g_R + g_N$ and $V = U_{N,N+1}$, we have
\begin{equation}
G_{N,N}^R = g_N + g_N\, U_{N,N+1}\, G^R_{N+1,N}
\end{equation}
and
\begin{equation}
G_{N+1,N}^R = g_R\, U_{N+1,N}\, G^R_{N,N}.
\end{equation}
(Note that the slice indices for the right Green's functions run
opposite to those in the left Green's functions.) Therefore,
\begin{equation}
G_{N,N}^R = \left( I - g_N\, U_{N,N+1}\, g_R\, U_{N+1,N} \right)^{-1}
g_N.
\end{equation}
Again, since $g_N = \left(E-h_N\right)^{-1}$, we can write
\begin{equation}
\label{eq:leadR}
G_{N,N}^R = \left( E - h_N - U_{N,N+1}\, g_R\,
U_{N+1,N} \right)^{-1}.
\end{equation}

The Green's function $G_{1,1}^L$ (or $G_{N,N}^R$) describes all
single-electron processes that begin and end that on the $n=1$ (or
$n=N$) slice, taking into account all possible number of incursions in
and out of the left (or right) lead. It does not yet take into account
incursions into the bulk of the system.

\subsection{Left Green's functions}
\label{sec:left}

With $G^L_{1,1}$ in hand, we can evaluate the next successive $N-1$
left Green's functions by using a recurrence formula analogous to
Eq. (\ref{eq:leadL}). To derive such formula, we choose $G^{(0)} =
G^L_{n-1,n-1}$ and $V = U_{n-1,n} + U_{n,n-1}$. Applying
Eq. (\ref{eq:dyson}), we write
\begin{equation}
G_{n,n}^L = \left( I - g_n\, U_{n,n-1}\, G_{n-1,n-1}^L\, U_{n-1,n}
\right)^{-1} g_n,
\end{equation}
with $n = 2,\ldots,N$. Using $g_n = \left(E - h_n \right)^{-1}$, we
obtain
\begin{equation}
\label{eq:GLrec}
 G_{n,n}^L = \left( E - h_n - U_{n,n-1}\,
G_{n-1,n-1}^L\, U_{n-1,n} \right)^{-1}.
\end{equation}
This formula is accompanied by another one, which connects the
left-most slice (the surface slice of the left lead) with the $n$th
one,
\begin{equation}
\label{eq:GLlong}
G_{0,n}^L = G^L_{0,n-1}\, U_{n-1,n}\,
G^L_{n,n}.
\end{equation}
Note that $N$ inversions are necessary to arrive at the $N$th
slice. Each inversion requires $O(M^3)$ operations. Thus, the
complexity of the calculation scales as $N\times M^3$.

\subsection{Right Green's functions}
\label{sec:right}

Similarly to left case, for the right Green's functions, using
Eq. (\ref{eq:leadR}) and starting from the $N$th slice, we find that
\begin{equation}
G_{n,n}^R = \left( I - g_n\, U_{n,n+1}\, G_{n+1,n+1}^R\, U_{n+1,n}
\right)^{-1} g_n,
\end{equation}
with $n = N-1,\ldots,1$. Substituting $g_n = \left(E -
h_n\right)^{-1}$, we obtain
\begin{equation}
\label{eq:GRrec}
 G_{n,n}^R = \left( E - h_n - U_{n,n+1}\,
G_{n+1,n+1}^R\, U_{n+1,n} \right)^{-1}.
\end{equation}
Also,
\begin{equation}
\label{eq:GRlong}
 G_{N+1,n}^R = G^R_{N+1,n+1}\, U_{n+1,n}\,
G^R_{n,n}.
\end{equation}
Again, $N$ additional inversions have to be performed in order to
arrive at slice the first slice ($n=1$), with an overall computation
cost $O(N\times M^3)$.

\subsection{Full Green's functions}
\label{sec:full}

Suppose one arrives at the $n$ slice by either a left or right sweep
($1 < n < N$). To obtain the exact full Green's function of the system
we use again Eq. (\ref{eq:dyson}) assuming $G^{(0)} = g_n +
G_{n-1,n-1}^L + G_{n+1,n+1}^R$, with $V=U_{n-1,n} + U_{n,n-1} +
U_{n,n+1} + U_{n+1,n}$. As a result, we find
\begin{equation}
G_{n,n} = g_n + g_n\, \left( U_{n,n-1}\, G_{n-1,n} + U_{n,n+1}\,
G_{n+1,n} \right),
\end{equation}
\begin{equation}
G_{n-1,n} = G^L_{n-1,n-1}\, U_{n-1,n}\, G_{n,n},
\end{equation}
and
\begin{equation}
G_{n+1,n} = G^R_{n+1,n+1}\, U_{n+1,n}\, G_{n,n}.
\end{equation}
Thus,
\begin{align}
G_{n,n} = \Big[ I - g_n\, ( U_{n,n-1}\, & G_{n-1,n-1}^L\,
U_{n-1,n} \nonumber\\
+&  U_{n,n+1}\, G_{n+1,n+1}^R\, U_{n+1,n} )\Big ]^{-1} g_n,
\end{align}
and since $g_n = \left(E - h_n \right)^{-1}$, we obtain
\begin{align}
\label{eq:Gexact}
G_{n,n} = \Big( E - h_n - \,& U_{n,n-1}\,G_{n-1,n-1}^L\, U_{n-1,n} 
\nonumber\\
-\,& U_{n,n+1}\, G_{n+1,n+1}^R\, U_{n+1,n}
\Big)^{-1},
\end{align}
together with
\begin{equation}
\label{eq:G0n}
 G_{0,n} = G^L_{0,n-1}\, U_{n-1,n}\, G_{n,n}
\end{equation}
and
\begin{equation}
\label{eq:GN+1n}
 G_{N+1,n} = G^R_{N+1,n+1}\, U_{n+1,n}\,
G_{n,n}.
\end{equation}
Note that in order to compute $G_{n,n}$ and $G_{N+1,n}$, we need to
keep track of $G^L_{n,n}$ and $G^R_{n,n}$ [obtained recursively from
Eqs. (\ref{eq:GLrec}) and (\ref{eq:GRrec}), respectively], as well as
$G^L_{0,n}$ and $G^R_{N+1,n}$ [which follow from
Eqs. (\ref{eq:GLlong}) and (\ref{eq:GRlong}), respectively]. In order
to obtain $G_{n-1,n}$ and $G_{n,n+1}$, we can apply Dyson's equation
again to a situation where only the $n$th slice is decoupled, yielding
\begin{equation}
\label{eq:Gnn+1}
G_{n,n+1} = G_{n,n}\, U_{n,n+1}\, G^R_{n+1,n+1},
\end{equation}
while
\begin{equation}
\label{eq:Gn-1n}
G_{n-1,n} = G^L_{n-1,n-1}\, U_{n-1,n}\,G_{n,n}.
\end{equation}
These equations are useful for computing the local current
distribution (Sec. \ref{sec:localquant}).

We note that when computing the exact Green's in
Eqs. (\ref{eq:Gexact}), (\ref{eq:G0n}), (\ref{eq:GN+1n}),
(\ref{eq:Gnn+1}), and (\ref{eq:Gn-1n}) we have selectively used each
matrix $U_{n,n^\prime}$ only once. Similarly, at each step, an
isolated slice Hamiltonian $h_n$ was used and never repeated. Thus, at
the end of the calculation of the full Green's function, all hoping
amplitudes and local potentials of the underlying tight-binding model
have been used and only once.

An alternative way to compute full Green's functions, which is quite
useful if only transmission and reflection matrices are required, is
to close the left (or right) sweep with a connection to the right
(left) lead:

\begin{enumerate}

\item For the left sweep, we use Eq. (\ref{eq:G0n}) to write
\begin{equation}
\label{eq:G0N+1}
G_{0,N+1} = G_{0,N}^L\, U_{N,N+1}\, G_{N+1,N+1},
\end{equation}
which is complemented by 
\begin{equation}
\label{eq:GN+1N+1}
G_{N+1,N+1} = \left(g_R^{-1} - U_{N+1,N}\, G^L_{N,N}\, U_{N,N+1}
\right)^{-1}
\end{equation}
obtained from Eq. (\ref{eq:Gexact}).

\item For the right sweep, we use instead Eqs. (\ref{eq:GN+1n}) and
(\ref{eq:G0n}) to obtain
\begin{equation}
\label{eq:GN+10}
G_{N+1,0} = G_{N+1,1}^R\, U_{1,0}\, G_{0,0}
\end{equation}
and
\begin{equation}
\label{eq:G00}
G_{0,0} = \left(g_L^{-1} - U_{0,1}\, G^R_{1,1}\, U_{1,0} \right)^{-1},
\end{equation}
respectively.

\end{enumerate}

As we will see below, Eqs. (\ref{eq:G0N+1}) and (\ref{eq:G00}) and can
be used to compute the left-to-right transmission and left reflection
matrices, respectively, while Eqs. (\ref{eq:GN+10}) and
(\ref{eq:GN+1N+1}) yield the right-to-left transmission and the right
reflection matrices. For systems with inversion symmetry, we expect
$G_{00} = G_{N+1,N+1}$ and $G_{0,N+1} = G_{N+1,0}$ and therefore only
one sweep (left or right) is necessary for the evaluation of the whole
scattering matrix.

For symmetric leads and in the absence of an external magnetic field
(i.e., time-reversal symmetric systems),
\be [G_{0,N+1}^r]^\dagger = G_{N+1,0}^a \ee
and only one sweep is necessary. When such conditions are not met, one
needs both sweeps, namely, from left-to-right and from right-to-left,
in order to assemble the scattering matrix. Moreover, any local
observable (such as the local density of states or the local current
flux), requires $G_{0,N+1}$ as well as $G_{n,n}$ for all
$n=1,\ldots,N$.

\subsection{Input Green's functions}
\label{sec:input}

The recurrence relations shown above rely on some input
information. One needs to define the Green's functions of the leads
($g_L$ and $g_R$), the Hamiltonian of the isolated slices ($h_n$,
$n=1,\ldots,N$), and the hopping between slices (the $U$ matrices)
before starting the calculation of the sample's Green's function.

Since both $g_L$ and $g_R$ are the input Green's functions, it is
crucial that they have finite imaginary parts. These will be dominant
and, in practice, we can basically neglect the imaginary part when
considering $g_n$ (even if $E$ happens to coincide with an eigenvalue
of an isolated slice, the imaginary parts brought in by coupling to
the leads makes the Green's function convergent). We will see next how
to obtain contact Green's functions for leads modeled as semi-infinite
lattices.

\section{Lead Green's Functions}
\label{sec:contacts}

To satisfactorily model the leads, there are two main physical
considerations to keep in mind: (a) the source and drain leads in
typical graphene transport experiments are metallic and thus have a
high density of states; (b) gra\-phene-metal interfaces tend to form
ohmic contacts. Thus, in the numerical simulations, one needs to
eliminate or minimize the contact resistance associated to band
structure mismatch at the contacts. To address (a), the chemical
potential in the leads is customarily adjusted to maximize the density
of states. To address (b), an appropriate lead lattice model,
compatible with the sample lattice, is chosen to minimize back
reflections. The leads are usually modeled either by square or
honeycomb semi-infinite lattices \cite{Schomerus07}. In certain cases,
a combination of square lattice contacts coupled to semi-infinite
linear chains are shown to be advantageous to minimize the contact
resistance \cite{Areshkin09}.

The methods to compute the lead Green's functions can be divided into
two categories, namely, the iterative recursive methods
\cite{LopezSancho85} and the eigenchannel decomposition or mode
matching ones \cite{MacKinnon85,Umerski97,Rocha06,Wimmer09}. In the
latter, the lead Green's function are built with the eigenchannels of
the {\it infinite} (translation invariant) corresponding
lattice. Except for few cases, such as the semi-infinite square
lattice discussed below, the eigenmodes depend on the longitudinal
wave number $k$ and the $g_{L,R}$ cannot be written in closed
analytical form.

In this Section we review the eigenchannel decomposition of $g_{L,R}$
for semi-infinite square lattices and discuss how to couple $g_{L,R}$
to a graphene device. Next, we present the decimation method for
semi-infinite lattices \cite{LopezSancho85}. Despite the claim that,
in general, iterative methods are inferior in performance and accuracy
than the eigendecomposition ones \cite{Umerski97}, the decimation
method has the attractive properties of being very robust and
straightforward, allowing for a very amenable implementation.

\subsection{Square lattice leads -- analytical approach}
\label{sec:analytic-squarelattice}

Let us model the contacts by semi-infinite tight-binding squa\-re
lattices \cite{Schomerus07}. Let us also set the Fermi energy to $E=0$
\cite{Tworzydlo06,Schomerus07}. We can shift the energy band of the
electronic states in the leads by varying a gate potential $V_{\rm
  lead}$ in the leads \cite{Rycerz07,Tworzydlo06,Schomerus07}. At zero
bias and in the absence of inelastic scattering, this sets the energy
of the electrons propagating through the graphene sheet.

The electron wave function in the semi-infinite square lattice is
extended along the $x$-direction and is quantized in the transverse
direction. Let us consider hard-wall boundary conditions at the edges
of the strip, namely, $j=0$ and $j=M+1$. The transverse wave
functions are given by
\be 
\chi_\nu(j) = \sqrt{\frac{2}{M+1}} \sin\left(\frac{\pi \nu
j}{M+1}\right),
\ee
where $\nu = 1, \cdots, M$. Associated to each transverse mode there
are two extended Bloch waves with longitudinal wave numbers $\pm k_n$,
which are real for propagating modes and complex for evanescent modes.

For convenience, we assume \cite{Ferry97} a hard-wall boundary
condition at the left end of the strip (where it connects to the right
Green's function of the strip), such that $\phi_\mu(N-1)=0$:
\be \phi_\mu(n)=\sqrt{\frac{2}{\pi}} \sin [\mu(n-N+1)], 
\ee
where $\mu$ is a longitudinal quantum number. When contrasted with a
continuum model, we can identify $\mu=k_x^\mu a_x$, where $k_x^\mu$ is
the longitudinal wave vector. By choosing $\phi_\mu(N-1)=0$ we are not
capable of treating the contribution of evanescent modes. This is
usually not a problem in practice, unless measurements are done very
close to the contacts (i.e., for very short systems).

The dispersion relation for this model is 
\be E_{\nu \mu} = V_{\rm lead} - 2t_x \cos\mu - 2 t_y \cos
\left(\frac{\pi \nu}{M+1}\right).  \ee
The velocities of the propagating modes are 
\begin{equation}
v_\nu = \frac{a_0}{\hbar} \left( \frac{dE_{\nu\mu}}{d\mu} \right) =
\frac{2a_0 t_x }{\hbar}\, \sin\mu.
\end{equation}
By setting $E_{\nu\mu}=0$, one writes $\sin\mu = \sqrt{1 - \cos^2\mu}$
with $\cos\mu = V_{\rm lead}/2t_x -(t_y/t_x) \cos \left[
  \pi\nu/(M+1)\right]$.

Let us now construct the leads Green's function as in
Ref. \cite{Ferry97}. The general expression
\be g(n,j;n^\prime,j^\prime;E) = \int_0^\pi d\mu \sum_{\nu=1}^M
\frac{[\phi_\mu(n)\chi_\nu(j)]^*[\phi_\mu(n^\prime)\chi_\nu(j^\prime)]}
     {E - E_{\nu \mu} + i0^+} \ee
can be simplified since we are only interest in $n=n^\prime=N$
(surface) and $E=0$. It reads
%
\be g_L(j,j^\prime) = \frac{2}{\pi} \sum_{\nu=1}^M
\chi_\nu(j)^*\chi_\nu(j^\prime) \int_0^\pi d\mu \frac{\sin^2 \mu} 
{p + q \cos\mu},  \ee
where 
\be p \equiv -V_{\rm lead} + 2t_y \cos\left(\frac{\pi
  \nu}{M+1}\right)+i0^+ \quad \mbox{and} \quad q\equiv 2 t_x. \ee
By integrating over $\mu$, one writes \cite{Ferry97} 
\be 
\label{eq:channel2site}
G^{\rm semi}_N(j,j^\prime) = \sum_{\nu=1}^M
\chi_\nu^*(j)\widetilde{G}^{\rm semi}(\nu)\chi_\nu(j^\prime) ,
\ee
where 
\be \widetilde{G}^{\rm semi}(\nu) = \frac{2p}{q^2}\left[ 1 -
\sqrt{1 - \left(\frac{q}{p}\right)^2} \right].
\ee
Equation (\ref{eq:channel2site}) defines the unitary transformation
that converts the Green's function $\widetilde{G}^{\rm semi}(\nu)$ in
the channel representation into $ G^{\rm semi}_N(j,j^\prime)$ in the
site representation.

In Ref. ~\cite{Schomerus07}, Schomerus argues that these expressions
can be related to the parameter $\mu$ for graphene semi-infinite
lattices in the case of armchair edge orientation. He shows that
$\widetilde{G}^{\rm semi}(\widetilde{\nu}) = - \mu^{\rm
  armchair}(V_{\rm lead})/t_x^2$. This correspondence works well for
armchair orientations because in that case propagating modes do not
mix; it does not work for zigzag or other edge orientations. In
Ref. \cite{Tworzydlo06} the authors speculate that transport
properties in the presence of bulk disorder should not depend on the
graphene orientation, which is numerically confirmed in
Ref.~\cite{Rycerz07}. Thus, in large-scale numerical simulations
involving bulk disorder, it is worth taking advantage of the matching
between square-lattice leads and graphene armchair leads.

\subsection{Eigenmode decomposition method -- square lattice}

The following alternative  approach to obtain the surface Gre\-en's
function of a square-lattice lead is helpful since the same steps can
be repeated for any other lattice and they form the basis of the
eigendecomposition methods. 

Since adding another slice to a semi-infinite lead should not alter
its Green's function, we can write that, in the absence of any
disorder or inhomogeneity,
\begin{equation}
G^L_{1,1} = g_L
\end{equation}
when $U_{0,1} = -t_x\, I$, where $t_x>0$ is the horizontal hopping
matrix element. Then, using Eq. (\ref{eq:leadL}), we obtain the
self-consistency condition
\begin{equation}
\label{eq:eigendecomp}
g_L = \left( E - h_1 - t_x^2\, g_L \right)^{-1}.
\end{equation}
Here, we assumed $U_{1,0}=t_x^2\, I$, appropriate for square lattices,
where $I$ is the identity matrix. In order to solve
Eq. (\ref{eq:eigendecomp}) for $g_L$, we notice that since $h_1$ is
Hermitian, it must be diagonalizable by a unitary transformation:
$T^\dagger\, h_1\, T = {\rm diag}(\varepsilon_\nu)$, where ${\rm
  diag}(\varepsilon_\nu)$ is a diagonal matrix containing the
eigenvalues of $h_1$. Then,
\begin{equation}
\label{eq:gLeq}
\tilde{g}_L (\nu) = \left( E - \varepsilon_\nu - t_x^2\, \tilde{g}_L
(\nu) \right)^{-1},
\end{equation}
where ${\rm diag} \left( \tilde{g}_L(\nu) \right) = T^\dagger\, g_L\,
T$ is the diagonal matrix containing the eigenvalues of $g_L$. Solving
Eq. (\ref{eq:gLeq}), we find
\begin{equation}
\tilde{g}_L(\nu) = \frac{(E-\varepsilon_\nu)}{2t_x^2} \pm
\sqrt{\frac{(E-\varepsilon_\nu)^2}{4t_x^4} - \frac{1}{t_x^2}}.
\end{equation}
Notice that for $|E-\varepsilon_\nu| < 2t_x$, this eigenvalue acquires
a finite imaginary part. Since we are mainly interested in retarded
Green's functions, we choose the negative sign and rewrite the
equation as
\begin{equation}
\tilde{g}_L^r(\nu) = \left\{ \begin{array}{ll}
  \frac{(E-\varepsilon_\nu)}{2t_x^2} \left[ 1 - \sqrt{1 -
      \frac{4t_x^2}{(E-\varepsilon_\nu)^2}} \right] + i0^+, & |E -
  \varepsilon_\nu| \ge 2t_x, \\ \frac{(E-\varepsilon_\nu)}{2t_x^2} -
  i\, \sqrt{\frac{1}{t_x^2} - \frac{(E-\varepsilon_\nu)^2}{4t_x^4}}, &
  |E - \varepsilon_\nu| < 2t_x. \end{array} \right.
\end{equation}

We need now to determine $T$ and $\{\varepsilon_\nu\}$. For a square
lattice, this is trivial: $h_1$ describes a one-dimensional chain with
$M$ sites and vertical hopping matrix elements $t_y>0$ (hard boundary
conditions assumed). Then,
\begin{equation}
T_{\nu j} = \sqrt{\frac{2}{M+1}}\, \sin \left( \frac{\pi\, \nu\,
  j}{M+1} \right) = \chi_\nu(j),
\end{equation}
where $j=1,\ldots, M$ and $\nu = 1, \ldots, M$, and
\begin{equation}
\varepsilon_\nu = V_{\rm lead} -2t_y\, \cos \left( \frac{\pi\,
  \nu}{M+1} \right),
\end{equation}
resulting in
\begin{equation}
\label{eq:gLsq}
g_L(j,j^\prime) = \sum_{\nu=1}^M \chi_\nu(j)\, \tilde{g}_L(\nu)\,
\chi_\nu(j^\prime).
\end{equation}
In Appendix \ref{sec:appendixA}, we show that Eq.~\eqref{eq:gLsq}
gives the expected steps in the linear conductance.

Unfortunately, the above expressions do not directly apply to either
zigzag or armchair leads, basically because $U_{0,1}$ is not
proportional to the identity in these cases and it does not commute
with the slice Hamiltonian. The solution for the general case is
nicely presented in Ref.~\cite{Wimmer09}.

\subsection{Leads Green's function -- decimation method}
\label{sec:decimation}

We can use the decimation method of Ref. \cite{LopezSancho85} to
evaluate numerically the Green's function of leads with arbitrary (but
translation invariant) lattice structures.

\begin{figure}[ht]
\centering
\includegraphics[width=0.95\columnwidth]{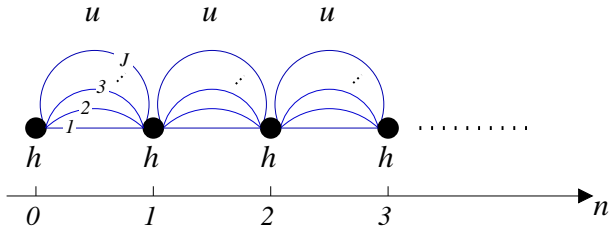}
\caption{Chain structure used in the decimation method.}
\label{fig:dec_chain}
\end{figure}

The method works as follows. Suppose we want to solve $(E-H)\, G(E) =
I$ for the operator $G(E)$, where the Hamiltonian operator $H$ is
defined over a chain where each site has an arbitrary basis of
dimension $J$ but only nearest-neighbor inter-site connections exist
(see Fig. \ref{fig:dec_chain}). We assume that all these connections
are represented by operators $u$ (left-to-right) and $u^\dagger$
(right-to-left). The local Hamiltonian is identical for all sites and
denoted by $h$. Since, by definition, the Green's function obeys
\begin{equation}
\left[ (E-H)\, G(E) \right]_{n,m} = \delta_{n,m},
\end{equation}
it follows that
\begin{eqnarray}
\label{eq:decG00}
(E-h)\, G_{0,0} & = & I + u\, G_{1,0}, \\
\label{eq:decG10}
(E-h)\, G_{1,0} & = & u\, G_{2,0} + u^\dagger G_{0,0}, \\ & \vdots &
\nonumber \\
\label{eq:decGn0}
(E-h)\, G_{n,0} & = & u\, G_{n+1,0} + u^\dagger G_{n-1,0},
\end{eqnarray}
with $n\ge 1$. Using Eq. (\ref{eq:decG10}), we find that
\begin{equation}
G_{1,0} = (E-h)^{-1} \left( u\, G_{2,0} + u^\dagger G_{0,0} \right),
\end{equation}
which can be combined with Eq. (\ref{eq:decG00}) to yield
\begin{equation}
\label{eq:decG00orig}
(E-h)\, G_{0,0} = I + u\, (E-h)^{-1} \left( u\, G_{2,0} + u^\dagger
G_{0,0} \right),
\end{equation}
which can be rewritten as
\begin{equation}
\label{eq:decG0}
\left[E-h - u\, (E-h)^{-1}u^\dagger \right]\,G_{0,0} = I + u\,
(E-h)^{-1} u\, G_{2,0}.
\end{equation}
Notice that we can relate $G_{2,0}$ to $G_{0,0}$ without involving
$G_{1,0}$.

The same trick can be employed for any value of $n$. From Eq.
(\ref{eq:decGn0}), we can write
\begin{equation}
G_{n+1,0} = (E-h)^{-1} \left( u\, G_{n+2,0} + u^\dagger G_{n,0}
\right),
\end{equation}
and
\begin{equation}
G_{n-1,0} = (E-h)^{-1} \left( u^\dagger G_{n-2,0} + u\, G_{n,0}
\right).
\end{equation}
These equations can be combined with Eq. \eqref{eq:decGn0} to yield
\begin{align}
\label{eq:decGn0orig}
(E-h)\, G_{n,0} = &\, u\, (E-h)^{-1} \left( u\, G_{n+2,0} + u^\dagger
G_{n,0} \right) + \nonumber\\ &\, u^\dagger (E-h)^{-1} \left( u\,
G_{n,0} + u^\dagger G_{n-2,0} \right),
\end{align}
which can be rewritten as
\begin{align}
\label{eq:decGn}
& \left[E-h - u\, (E-h)^{-1} u^\dagger - u^\dagger (E-h)^{-1} u
  \right]\, G_{n,0} = \nonumber\\ & \quad \qquad u\, (E-h)^{-1} u\,
G_{n+2,0} + u^\dagger (E-h)^{-1} u^\dagger G_{n-2,0}.
\end{align}

Equations (\ref{eq:decG0}) and (\ref{eq:decGn}) generate a new
recursion series involving only even sites:
\begin{eqnarray}
\label{eq:dec_rec0}
(E-\varepsilon_1^s)\, G_{0,0} & = & I + \alpha_1\, G_{2,0} \\
(E-\varepsilon_1)\, G_{2,0} & = & \alpha_1\, G_{4,0} + \beta_1\,
G_{0,0} \\ & \vdots & \nonumber \\
\label{eq:dec_recn}
(E-\varepsilon_1)\, G_{n,0} & = & \alpha_1\, G_{n+2,0} + \beta_1\,
G_{n-2,0}
\end{eqnarray}
with
\begin{eqnarray}
\alpha_1 & = & u\, (E-h)^{-1} u \\ \beta_1 & = & u^\dagger (E-h)^{-1}
u^\dagger \\ \varepsilon_1^s & = & h + u\, (E-h)^{-1}u^\dagger
\\ \varepsilon_1 & = & \varepsilon_1^s + u^\dagger (E-h)^{-1} u.
\end{eqnarray}
Even though Eqs. (\ref{eq:dec_rec0}) to (\ref{eq:dec_recn}) involve
only even sites (i.e., multiples of $2^1$), they are identical in form
to Eqs. (\ref{eq:decG10}) and (\ref{eq:decGn0}). Therefore, we can
repeat this procedure $k$ times until the recursion relations involve
only sites that are multiple of $2^k$, namely,
\begin{eqnarray}
(E-\varepsilon_k^s)\, G_{0,0} & = & I + \alpha_k\, G_{2,0} \\
(E-\varepsilon_k)\, G_{2^k,0} & = & \alpha_k\, G_{2^k \cdot 2,0} +
\beta_k\, G_{0,0} \\ & \vdots & \nonumber \\ (E-\varepsilon_k)\,
G_{2^k \cdot n,0} & =&  \alpha_k\, G_{2^k \cdot (n+1),0} + \beta_k\,
G_{2^k \cdot (n-1),0}
\end{eqnarray}
with $n\ge 1$ and
\begin{eqnarray}
\alpha_k & = & \alpha_{k-1}\, (E- \varepsilon_{k-1})^{-1} \alpha_{k-1},
\\ \beta_k & = & \beta_{k-1}\, (E- \varepsilon_{k-1})^{-1} \beta_{k-1},
\\ \varepsilon_k^s & = & \varepsilon_{k-1} + \alpha_{k-1} (E-
\varepsilon_{k-1})^{-1}\, \beta_{k-1}, \\ \varepsilon_k & = &
\varepsilon_k^s + \beta_{k-1} (E- \varepsilon_{k-1})^{-1}\alpha_{k-1}.
\end{eqnarray}

The decimation can stop when $||\alpha_k||$ and $||\beta_k||$ are
sufficiently small, in which case we can approximate
\begin{equation}
G_{0,0} \approx (E- \varepsilon_k^s)^{-1}.
\end{equation}
This provides the Green's function for the ``surface'' slice, which
can then be related to the lead Green's functions $G^L_0$ and
$G_{N+1}^R$.

Note that one needs to add a small positive imaginary part to $E$,
namely, $E \rightarrow E + i\eta$, in order to generate {\it retarded}
Green's functions. On the practical side, an increase of $\eta$ helps
to speed up the convergence but spoils the accuracy of $G^r_{0,0}$ at
the order of $\eta/t$. Provided that $\eta/t \ll 1$, adding an
imaginary part to $E$ has little effect on the computation of graphene
transport properties for $|E|/\eta \gg 1$, except for a small energy
interval around the charge neutrality point, where $E=0$.

\subsubsection{Square lattice lead}
\label{sec:squarelead}

\begin{figure}[t]
\centering
\includegraphics[width=0.8\columnwidth]{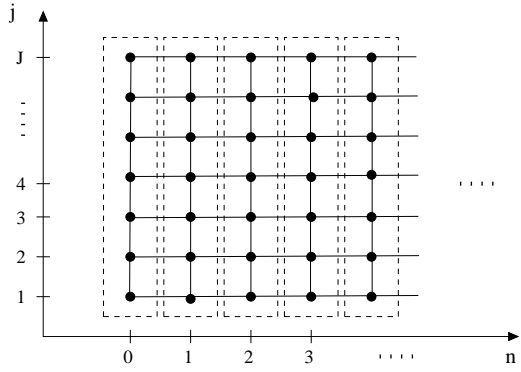}
\caption{Structure of the square lattice lead. The slice with $n=0$
  corresponds to the ``surface'' which will be attached to the
  system. Slices for decimation are denoted by the boxes with dashed
  lines.}
\label{fig:squarelead}
\end{figure}

From Fig. \ref{fig:squarelead}, we see that each decimation site
corresponds to a regular open vertical chain with $J$ lattice
sites. Therefore,
\begin{equation}
\label{eq:h}
h(j,j^\prime) = \sum_{\mu=1}^J \phi_\mu(j)\, \phi_\mu(j^\prime)\, E_\mu,
\end{equation}
and
\begin{equation}
\label{eq:u}
u(j,j^\prime) = -t\, \delta_{j,j^\prime}.
\end{equation}
where $t$ is the nearest-neighbor hopping amplitude, $j,j^\prime =
1,\ldots, J$, $\mu = 1, \ldots, J$,
\begin{align}
\phi_\mu(j) =&\, \sqrt{\frac{2}{J+1}}\, \sin \left( \frac{\pi\mu j}{J+1}
\right), \quad {\rm and} \quad \nonumber\\
E_\mu = &\,-2t\, \cos \left(
\frac{\pi\mu}{J+1} \right).
\end{align}
From Eqs. (\ref{eq:h}) and (\ref{eq:u}), we derive the relations
\begin{equation}
\alpha_1(j,j^\prime) = t^2\, \left[ (E-h)^{-1} \right]_{j,j^\prime},
\end{equation}
\begin{equation}
\varepsilon_1^s (j,j^\prime) = h(j,j^\prime) + \alpha_1(j,j^\prime),
\end{equation}
and
\begin{equation}
\varepsilon_1^s (j,j^\prime) = h(j,j^\prime) + 2 \alpha_1(j,j^\prime).
\end{equation}

Figure \ref{fig:G_sp_leads} shows a comparison between the analytical
and decimation results for the $G^r(E)$ of a square lattice of width
$J=M=6$, projected onto the eigenchannel basis. Notice the excellent
agreement.
 
\begin{figure}[ht]
\centering
\includegraphics[width=0.95\columnwidth]{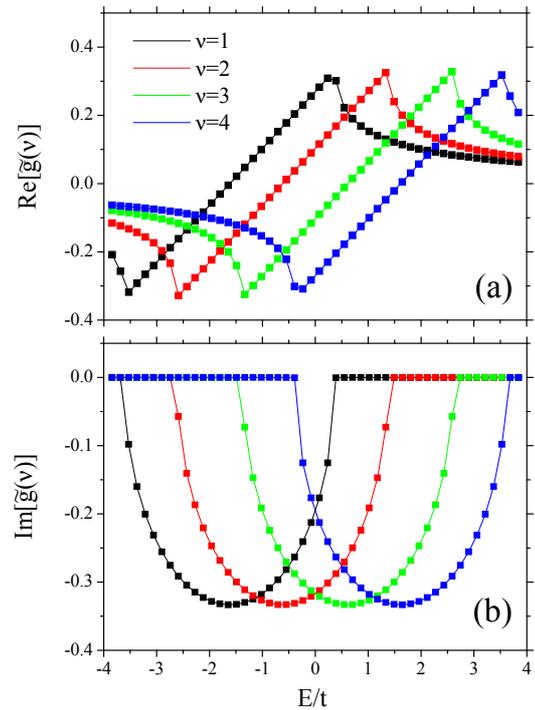}
\caption{
Real (a) and imaginary part (b) of the Green function of a
semi-infinite square lattice in the channel (diagonal) representation,
$\widetilde{g}(\mu)$, as a function of energy. $J=4$. The solid lines
represent the exact analytic result, whereas the squares were 
numerically obtained through the decimation method for $\eta=10^{-6}$ and 
$k=10$.
}
\label{fig:G_sp_leads}
\end{figure}

\subsubsection{Honeycomb lattice lead -- armchair edges}
\label{sec:armchairlead}

\begin{figure}[ht]
\centering \includegraphics[width=0.95\columnwidth]{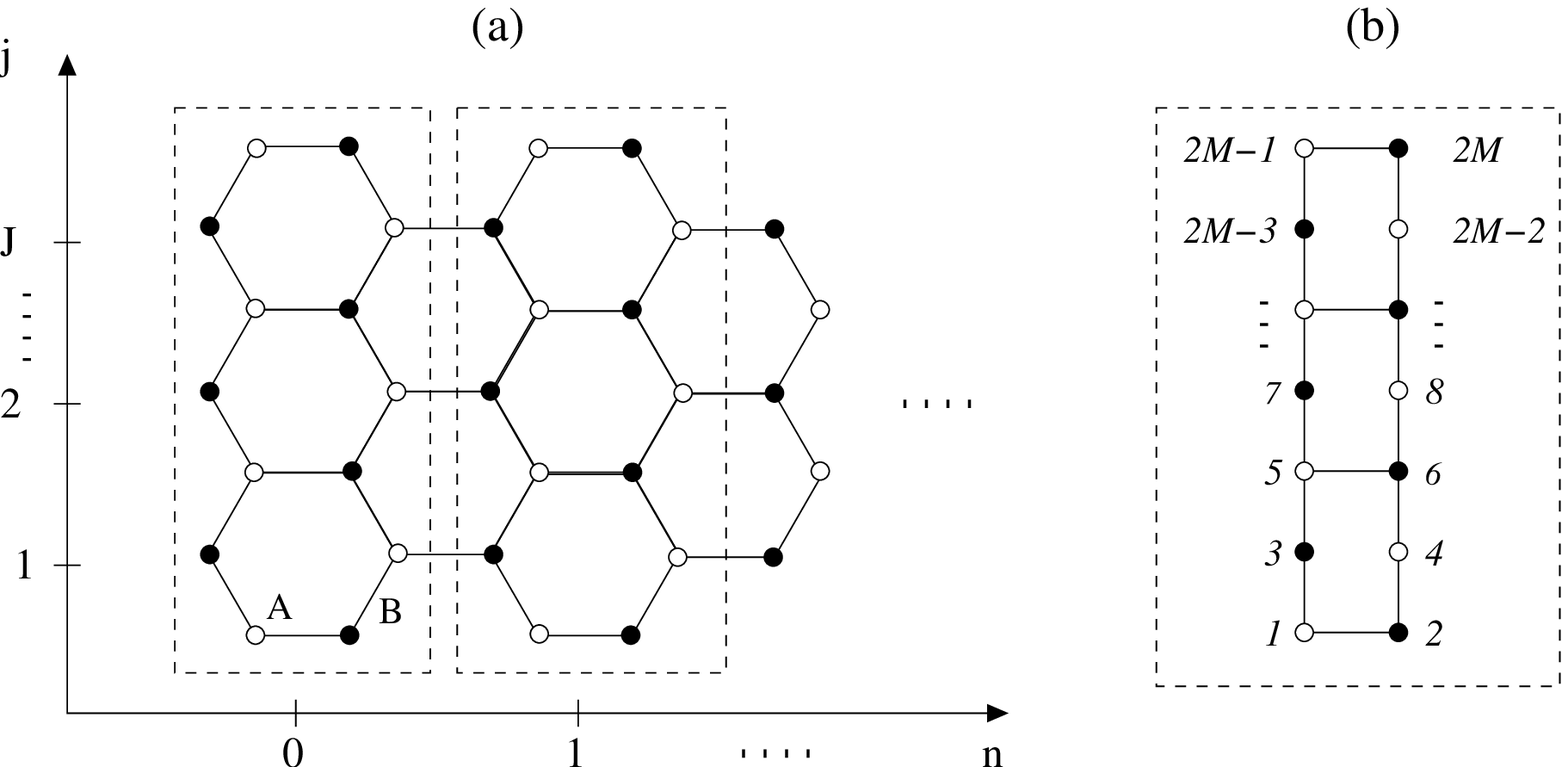}
\caption{(a) Structure of the honeycomb armchair lead. Note that the
  number of interconnect channels runs from 1 to $J$, while the number
  of vertical coordinate points in the slice is $M=2J+1$. The slices
  for decimation are defined by the boxed regions. (b) Internal index
  structure of the elementary slice.}
\label{fig:armchairlead}
\end{figure}

For the honeycomb lattice with armchair edges, the elementary slice
contains $J$ stacked hexagons. Thus, $M=2J+1$, where $M$ is the
vertical number of atoms. The slice has the structure of a vertical
ladder chain ($2M$ atoms). In this case there is no simple formula for
the eigenstates of $h$. The matrix $h$, which is $2M\times 2M$
dimensional, reads
\begin{equation}
\tilde{h} = \left( \begin{array}{ccccccccc} 
& -t & -t & & & & & & \\
-t & & & -t & & & & & \\ 
-t & & & -t & -t & & & & \\ 
& -t & -t & & & -t & & & \\ 
& & -t & & & -t & -t & & \\ 
& & & -t & -t & & & -t & \\ 
& & & & -t & & & -t & \\ 
& & & & & -t & -t & & \\
& & & & & & & & \ddots 
\end{array} \right).
\end{equation}
Let us call $\{\phi_\mu(m)\}$ its eigenvectors and $\{E_\mu\}$ its
eigenvalues, with $\mu=1,\ldots, 2M$. Then,
\begin{equation}
h(j,j^\prime) = \sum_{\mu=1}^{2M} \phi_\mu(4j)\,
\phi_\mu(4j^\prime-1)\, E_\mu,
\end{equation}
with $j,j^\prime = 1, \ldots, J$. Everything else is similar to
Sec. \ref{sec:squarelead}.

\subsubsection{Honeycomb lattice lead -- zigzag edges}
\label{sec:zigzaglead}

\begin{figure}[t]
\centering \includegraphics[width=0.95\columnwidth]{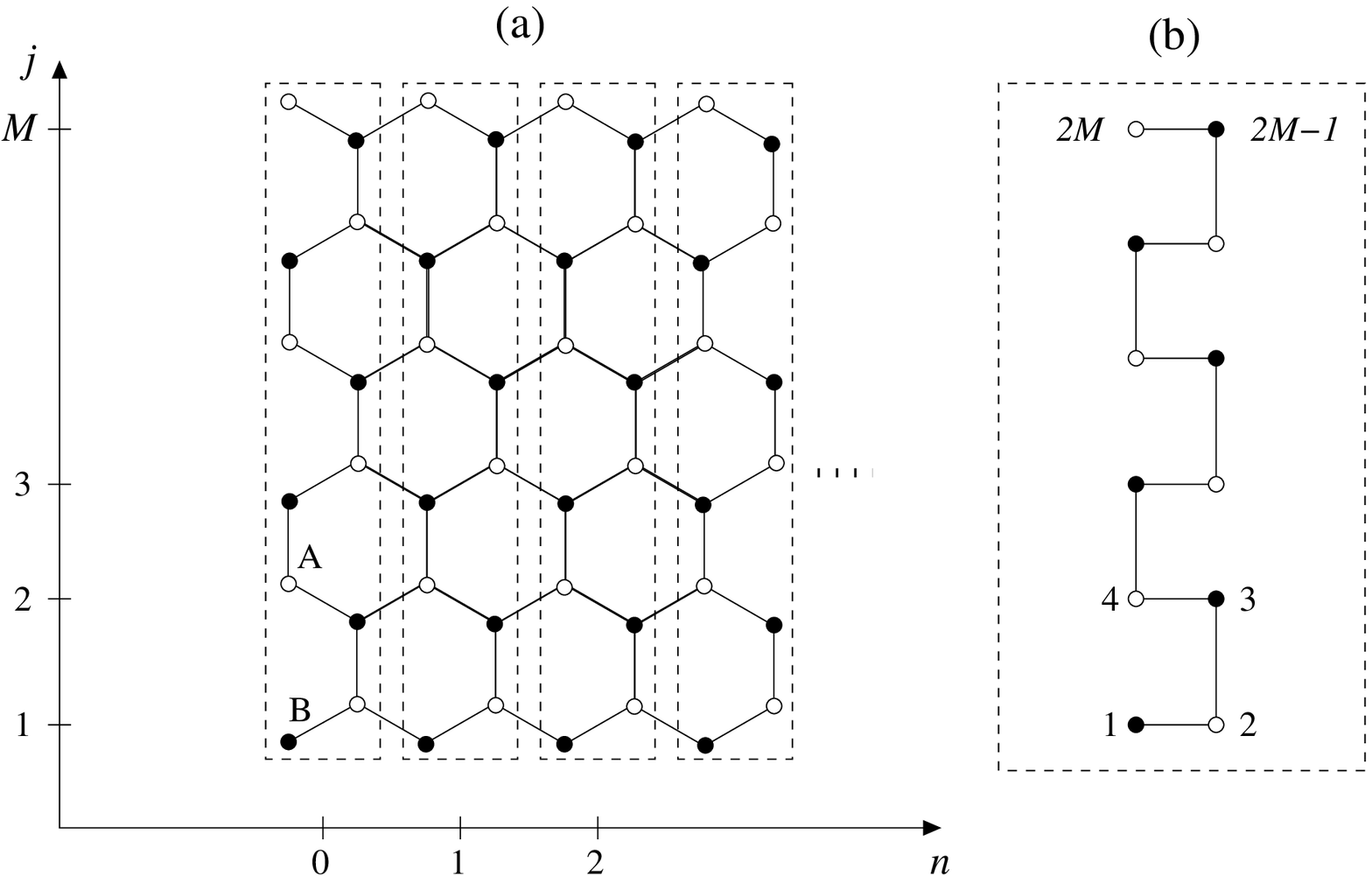}
\caption{(a) Structure of the honeycomb zigzag lead. Note that the
  number of interconnect channels runs from 1 to $J=M$, where $M$ is
  the number of vertical coordinate points in the graphene. The slices
  for decimation are defined by the boxed regions. (b) Internal index
  structure of the elementary slice.}
\label{fig:zigzaglead}
\end{figure}

For the zigzag graphene lead, there is a direct correspondence between
slice channels and the vertical indices of atomic positions:
$M=J$. The elementary slice is a open vertical chain of length
$2M$. Its eigenfunctions and eigenvalues are
\begin{align}
&\phi_\mu(l) = \sqrt{\frac{2}{2M+1}}\, \sin \left( \frac{\pi\mu
l}{2M+1} \right), \quad {\rm and} \nonumber\\ 
& E_\mu =  -2t\, \cos \left(\frac{\pi\mu}{2M+1} \right),
\end{align}
with $l = 1,\ldots, M$ and $\mu = 1, \ldots, 2M$. The slice matrix
reads
\begin{equation}
h(j,j^\prime) = \sum_{\mu=1}^{2M} \phi_\mu(2j-1)\,
\phi_\mu(2j^\prime)\, E_\mu,
\end{equation}
with $j,j^\prime = 1, \ldots, M$. All other aspects are identical to
Sec. \ref{sec:squarelead}.

\section{Device Green's Function}
\label{sec:tightbind}

The device Green's function $G_{i,j}$ can be used to describe nearly
all the physics of transport and its calculation is where most of the
computational time is spent. This is where any optimization of the
computational method is most welcome, particularly in the study of
disorder effects in the electronic transport when extensive disorder
averaging is required. With the RGF method one can compute $G_{i,j}$
for a large variety of settings. For instance, the graphene sheet can
have an arbitrary number of layers and different edge orientations,
namely, zigzag, armchair, and chiral. Also, the tight-binding model
can include next-nearest neighbor hopping terms in addition to the
nearest neighbor ones. These elements have to be taken into account
when choosing the slice unit cell employed by the recursive method. The
description of disorder modeling is postponed to
section~\ref{sec:disorder}.

In this Section we present efficient slicing schemes for graphene
monolayers with armchair and zigzag edges.

\subsection{Slicing armchair lattices}
\label{sec:armchair_eff}

For rectangular geometries, the bottleneck of the recursive method is
the matrix inversion required for adding a new slice [see
  Eq. (\ref{eq:GLrec})]. Thus, it is always important to try to
minimize the number of sites in the slice. Having this in mind, there
is a way to mount the armchair slices which reduces the number of
sites per slice without introducing next-to-nearest neighbor
hopping. It is based on the lattice deformation shown in
Fig. \ref{fig:armchair_eff}.

\begin{figure}[ht]
\centering
\includegraphics[width=0.85\columnwidth]{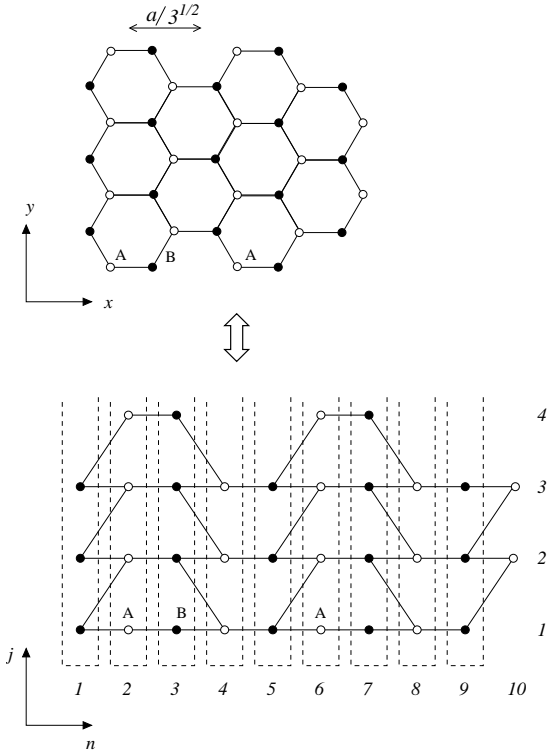}
\caption{Efficient way to slice an armchair graphene ribbon (the
  so-called ``pine tree" configuration). Atoms from different
  sublattices are indicated by empty (A) and full (B) circles. $a_0$
  is the lattice constant. The dashed lines indicate vertical slices.}
\label{fig:armchair_eff}
\end{figure}

The relations between lattice size and graphene sheet dimensions for
the efficient armchair slicing are
\begin{equation}
\frac{L}{a_0} = \frac{\sqrt{3}}{2} \left( \frac{N}{2} - 1 \right) +
\frac{\sqrt{3}}{6} \quad {\rm and} \quad \frac{W}{a_0} = M-1.
\end{equation}
%

\subsection{Slicing zigzag lattices}
\label{sec:zigzag}

The zigzag geometry is shown in Fig. \ref{fig:zigzag}. The real
structure is shown on the left-hand side (honeycomb lattice). An
equivalent square lattice with missing vertical bonds is also shown
(the so-called ``brick wall'' configuration). For zigzag edge ribbons
there is no efficient, alternative slicing that minimizes the number
of sites per slice {\it without} creating a next-to-nearest neighbor
connectivity.

\begin{figure}[ht]
\centering
\includegraphics[width=0.75\columnwidth]{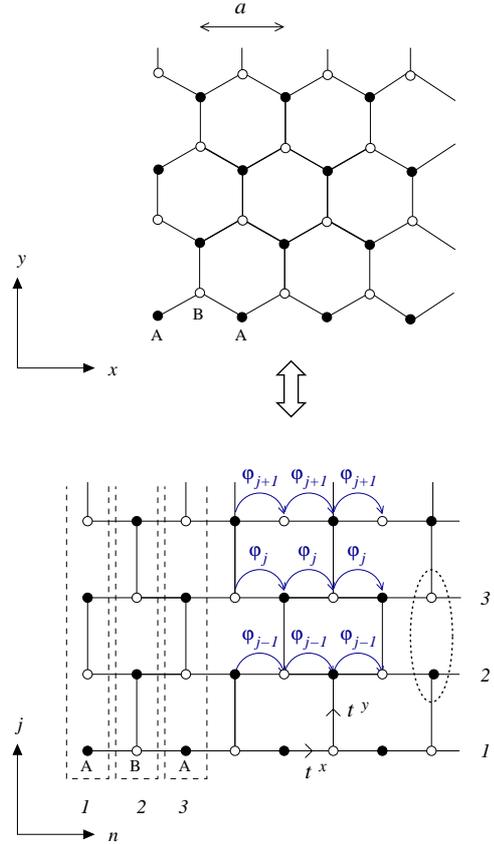}
\caption{Graphene strip with zigzag edges. Atoms from different
  sublattices are indicated by empty (A) and full (B) circles. $a_0$
  is the lattice constant. The dashed lines indicate vertical slices
  and the dotted highlights a dimer. By convention, we set as (1,1)
  the coordinates of an atom of type A placed at the left bottom
  corner. The Peierls phases (see appendix \ref{sec:A}) of the
  ``horizontal'' hopping matrix elements are indicated.}
\label{fig:zigzag}
\end{figure}

By convention, we assume that the site at the left bottom corner of
the lattice is of type A. In that way, slices with even and odd number
of sites will alternate as we move horizontally. Two cases will need
to be considered separately: $M$ odd and $M$ even. However, both cases
share a common trend, namely, the presence of dimers and isolated
sites (at the bottom and/or at the top of the slice). Therefore,
finding the Green's function of an isolated slice is a very simple
exercise.

The relations between lattice size and graphene sheet dimensions for
the zigzag slicing are
\begin{equation}
\frac{W}{a_0} = \frac{\sqrt{3}}{2} (M-1) + \frac{\sqrt{3}}{6} \quad
{\rm and}  \quad \frac{L}{a_0} = \frac{N-1}{2}.
\end{equation}
%

\section{Evaluating Local Quantities}
\label{sec:localquant}

In addition to the transmission ${\cal T}$, the RGF method can be used
to calculate other quantities such as the local density of states
(e.g. \cite{Metalidis05}) and the local current density
(e.g. \cite{Todorov02,Cresti03}).

\subsection{Local density of states}
\label{sec:LDOS}

The local density of states (LDOS) can be easily evaluated if one knows
the exact retarded Green's function at a given site:
\begin{equation}
\rho(n,j;E) = -\frac{1}{\pi}\, {\rm Im} \left[ G^r_{n,n}(j,j;E) \right].
\end{equation}
The exact local Green's function is evaluated using
Eq.~(\ref{eq:Gexact}). Conductance calculations require just a single
sweep through the lattice, since the transmission formula
\eqref{eq:caroli} only needs $G^r$ evaluated at the slices
corresponding to the contacts. In contrast, the LDOS demands the
calculation of $G^r$ at all lattice slices of interest, which further
increases the computational cost linearly with $N$, resulting in
$O(N^2M^3)$. Let us provide an example of a situation where the LDOS
plays a central role and needs to be evaluated.

Numerous studies have addressed the possibility of local magnetic
moment formation in graphene either due to zigzag terminations at the
edges in graphene nanoribbons or due to vacancies in graphene sheets
(for a review, see Ref.~\cite{Yazyev10}). The effect can be understood
through the Stoner mechanism of magnetism, which requires an enhanced
LDOS when strong electron-electron interaction as present, as
originally proposed in Ref. \cite{Son06} using the Density Functional
Theory (DFT). The tight-binding Hamiltonian,
Eq.~\eqref{eq:tight-bindingH}, can be modified to reproduce the DFT
results by adding a Hubbard mean-field term \cite{Son06}, namely,
\begin{align}
\label{eq:HMF}
H = -\sum_{i<j,\sigma} &\left( t_{ij} c^\dagger_{i, \sigma} c^{}_{j,
  \sigma} + {\rm H. c.} \right) \nonumber\\ &+\,\sum_{i, \sigma}\Big(
V_i + U \langle n_{i,-\sigma} \rangle \Big) c^\dagger_{i, \sigma}
c^{}_{i,\sigma},
\end{align}
where the operators $c^\dagger_{i,\sigma}$ and $c^{}_{i,\sigma}$
create and annihilate an electron of spin projection $\sigma$ at at
the site $i$. As standard, $\langle n_{i,\sigma} \rangle$ is the
occupation number and $U$ is the on-site electron-electron interaction
strength. In graphene, $U$ is usually fitted to reproduce the DFT band
structure calculations for translation invariant systems (for which
$V_i$ is constant). As we discuss in Sec.~\ref{sec:disorder}, by a
suitable choice of the parameters $V_i$ and $t_{ij}$, the Hamiltonian
\eqref{eq:HMF} becomes an excellent framework to model disorder as
well.

In general, the occupation numbers $\langle n_{i,\sigma} \rangle$ that
appear in Eq. (\ref{eq:HMF}) can be obtained from non\-equilibrium
Green's functions \cite{HaugJauho08}. In the linear response regime, a
simplification allows one to express $\langle n_{i,\sigma} \rangle$ in
terms of an equilibrium Green's function, namely,
\be
\label{eq:occupation}
\langle n_{i,\sigma} \rangle = - \frac{1}{\pi} \int_{-\infty}^{\infty}
dE \,\mbox{Im}\,[G^r_{n,n;\sigma}(j,j;E)] f(E - \mu).  \ee
where $i\equiv(j,n)$. Note that Eqs. \eqref{eq:HMF} and
\eqref{eq:occupation} have to be solved self-consistently.

The large number of poles of $G^r(E)$ makes impractical the
integration of the {\it r.h.s.} of Eq.~\eqref{eq:occupation} on the
real axis, a difficulty shared with transport studies on molecular
electronics, see e.g. Ref. \cite{Xue01}. An optimized strategy to
implement an efficient integration along a complex plane contour is
presented in Refs.~\cite{Areshkin09,Areshkin10}.

Another frequent application of LDOS occurs in the evaluation of the
local charge density. The latter can be calculated from the LDOS
through the expression
\begin{equation}
n_c(n,j;E_F) = \frac{1}{A}\, \int_{\mu(n,j)}^{E_F} dE\, \rho(n,j;E),
\end{equation}
where $A$ is the sheet area and $\mu(n,j)$ denotes the local chemical
potential. Note that in $n$-type regions, $E_F > \mu$ and therefore
the integral is over positive energies (``electrons''), while in
$p$-type regions, $E_F < \mu$ and the integral is over negative
energies (``holes''). The local chemical potential is evaluated with
respect to $E_F=0$,
\begin{equation}
\mu(n,j) = V(n,j),
\end{equation}
where the background potential $V(n,j)$ includes the gate voltage.

\subsection{Local current density}
\label{sec:currentdensity}

The local current density is involves more complex calculation than
the LDOS. Several methods were developed in the literature (see
e.g. Ref. \cite{Todorov02}). The bond current between two neighboring
sites of lattice coordinates $(n,j)$ and $(n^\prime,j^\prime)$ is
obtained using the equations-of-motion method for nonequilibrium
Green's functions (see e.g. \cite{HaugJauho08}) and reads
\cite{Caroli71}
\begin{align}
I_{(n,j)\rightarrow(n^\prime,j^\prime)} = -\frac{2e}{h} \int dE
&\left[\, U_{n,n^\prime}(j,j^\prime)
  G^<_{n^\prime,n}(j^\prime,j)\right.\nonumber\\&\left. \;\;\;\;-
  U_{n^\prime,n}(j^\prime,j) G^<_{n,n^\prime}(j,j^\prime) \right],
\label{eq:local_current}
\end{align}
where $G^<$ is the lesser Green's function of the
system. \footnote{There are several good textbooks, such as
  Refs.~\cite{HaugJauho08,BruusFlensberg04,NazarovBlanter09}, that
  discuss nonequilibrium Green's functions. In particular,
  Ref.~\cite{HaugJauho08} concisely covers all required background
  material. We refer the reader to these books for the derivation of
  expressions involving $G^<$ and related functions and further
  insight into the subject.}  For the tight-binding model with
nearest-neighbor hopping, there are two situations to consider. First,
$n=n^\prime$, in which case $j^\prime = j \pm 1$ and the current is
intra-slice. Second, when $n^\prime = n \pm 1$, the current is
inter-slice and $j^\prime = j$ or $j^\prime = j \pm 1$, at
most. Notice that in the absence of magnetic fields, time-reversal
symmetry requires $U_{n,n^\prime}(j,j^\prime) =
U_{n^\prime,n}(j^\prime,j)$. In addition, in equilibrium conditions,
the lesser Green's function is a symmetric matrix, leading to a zero
bond current, as expected. Thus, local currents can only appear
through the application of a magnetic field (which breaks
time-reversal symmetry) or when a finite bias voltage between contacts
exists.

Equation (\ref{eq:local_current}) requires the calculation of the
exact lesser Green's function for sites in the bulk of the
system. This can be done recursively, starting from the equilibrium
Green's function of the leads (see Sec. \ref{sec:Glesser}). An
alternative approach, suitable for transport in the linear regime, is
to use the elimination technique developed in Ref. \cite{Cresti03},
where only retarded and advanced Green's functions are required and
the energy integration is avoided. The computational cost is this
approach also scales as $O(N^2M^3)$.

For plotting current fields, it is useful to define the local,
on-site, outgoing {\it vector} current as
\begin{equation}
\label{eq:Idensity}
\vec{I}_{n,j} = \sum_{n^\prime,j^\prime} \vec{a}_{(n,j)\rightarrow
  (n^\prime,j^\prime)}\, I_{(n,j)\rightarrow (n^\prime,j^\prime)},
\end{equation}
where the sum is over sites $(n^\prime,j^\prime)$ that are nearest
neighbors to site $(n,j)$ and $\vec{a}_{k,k^\prime}$ is the lattice
vector between sites $k$ and $k^\prime$. Notice that this local vector
current does not necessarily fall along any of the bonds coming out of
the site $(n,j)$.

\subsection{Recursion for non-equilibrium Green's functions}
\label{sec:Glesser}

In order to evaluate the local current distribution
\eqref{eq:local_current} in the most general case, one needs to
determine the exact lesser Green's function, $G^<$. The procedure is
the following.

\begin{enumerate}

\item We start with the retarded Green's functions of the leads (which
  are assumed to be in equilibrium) and use the
  fluctuation-dissipation relations
\begin{align}
& g^<_L = -if_L \left( g_L^r - g_L^a \right), \quad \\
\mbox{and} \nonumber\\
& g^<_R = -if_R \left( g_R^r - g_R^a \right),
\end{align}
where the advanced Green's functions obey $g_L^a = \left( g_L^r
\right)^\dagger$ and $g_R^a = \left( g_R^r \right)^\dagger$. Here,
$f_L$ and $f_R$ denote the Fermi distributions in the left and right
leads, respectively.

\item For the left-to-right sweep, we first determine the retarded
  Green's function at the $n$th slice and then obtain the lesser
  Green's function using the expression
\begin{equation}
G^{L,<}_{n,n} = G^{L,r}_{n,n}\, \Sigma^{L,<}_{n,n}\, G^{L,a}_{n,n},
\end{equation}
where the self energy due to the coupling of the $n$th slice to all
other slices to the left is given by
\begin{equation}
\label{eq:SigmaL}
\Sigma^{L,<}_{n,n} = U_{n,n-1}\, G^{L,<}_{n-1,n-1}\, U_{n-1,n}.
\end{equation}
Notice that $G^{L,a}_{n,n} = \left( G^{L,r}_{n,n} \right)^\dagger$ and
$U_{n,n-1} = \left( U_{n-1,n} \right)^\dagger$. Thus, we obtain the
recurrence relation
\begin{equation}
G_{n,n}^{L,<} = \left( G_{n,n}^{L,r}\, U_{n,n-1}
  \right)\, G_{n-1,n-1}^{L,<}\, \left( G_{n,n}^{L,r}\, U_{n,n-1}
  \right)^\dagger.
\end{equation}

\item For the right-to-left sweep, we apply instead the analogous
  expression
\begin{equation}
G^{R,<}_{n,n} = G^{R,r}_{n,n}\, \Sigma^{R,<}_{n,n}\, G^{R,a}_{n,n},
\end{equation}
where
\begin{equation}
\label{eq:SigmaR}
\Sigma^{R,<}_{n,n} = U_{n,n+1}\, G^{R,<}_{n+1,n+1}\, U_{n+1,n}.
\end{equation}
Thus, we obtain the other recurrence relation
\begin{equation}
 G_{n,n}^{R,<} = \left( G_{n,n}^{R,r}\, U_{n,n+1}
  \right)\, G_{n+1,n+1}^{R,<}\, \left( G_{n,n}^{R,r}\, U_{n,n+1}
  \right)^\dagger.
\end{equation}

\item In order to join the two sweeps to obtain the exact lesser
  Green's function at a given slice, we simply combine left and right
  self energies and evaluate
\begin{equation}
G^<_{n,n} = G^r_{n,n} \left( \Sigma^{L,<}_{n,n} + \Sigma^{R,<}_{n,n}
\right) G^a_{n,n}
\end{equation}
using Eqs. (\ref{eq:SigmaL}) and (\ref{eq:SigmaR}) for the self
energies.

\item Once the left-to-right and right-to-left Green's functions are
  know, the exact inter-slice lesser Green's functions are obtained
  using the Dyson-Langreth equations \cite{HaugJauho08}
\begin{equation}
\label{eq:DL1}
G^<_{n-1,n} = G^{L,r}_{n-1,n-1}\, U_{n-1,n}\, G^<_{n,n} +
G^{L,<}_{n-1,n-1}\, U_{n-1,n}\, G^a_{n,n},
\end{equation}
\begin{equation}
\label{eq:DL2}
G^<_{n,n+1} = G^r_{n,n}\, U_{n,n+1}\, G^{R,<}_{n+1,n+1} + G^<_{n,n}\,
U_{n,n+1}\, G^{R,a}_{n+1,n+1},
\end{equation}
\begin{equation}
\label{eq:DL3}
G^<_{n,n-1} = G^r_{n,n}\, U_{n,n-1}\, G^{L,<}_{n-1,n-1} + G^<_{n,n}\,
U_{n,n-1}\, G^{L,a}_{n-1,n-1},
\end{equation}
and
\begin{equation}
\label{eq:DL4}
G^<_{n+1,n} = G^{R,r}_{n+1,n+1}\, U_{n+1,n}\, G^<_{n,n} +
G^{R,<}_{n+1,n+1}\, U_{n+1,n}\, G^a_{n,n}.
\end{equation}

\end{enumerate}

\subsection{Current conservation}

Notice that the current through consecutive slices is conserved.  Let
us prove that using Eq. \ref{eq:local_current} and writing
\begin{equation}
I_{(n-1)\rightarrow n} = - \frac{e}{h} \int dE\, {\rm Tr} \left(
G^<_{n,n-1}\, U_{n-1,n} - U_{n,n-1}\, G^<_{n-1,n} \right)
\label{eq:I(n-1)(n)}
\end{equation}
and
\begin{equation}
I_{n\rightarrow(n+1)} = - \frac{e}{h} \int dE\, {\rm Tr} \left(
G^<_{n+1,n}\, U_{n,n+1} - U_{n+1,n}\, G^<_{n,n+1} \right),
\label{eq:I(n)(n+1)}
\end{equation}
where the traces indicate a sum over all sites in the slices $n$ and
$n+1$, respectively. Since
\begin{equation}
G^<_{n,n-1}\, U_{n-1,n} = G^r_{n,n}\, \Sigma_{n,n}^{L,<} + G^<_{n,n}\,
\Sigma^{L,a}_{n,n},
\end{equation}
\begin{equation}
U_{n,n-1} \, G^<_{n-1,n} = \Sigma_{n,n}^{L,r}\, G^<_{n,n} +
\Sigma^{L,<}_{n,n}\, G^a_{n,n},
\end{equation}
\begin{equation}
G^<_{n,n+1}\, U_{n+1,n} = G^r_{n,n}\, \Sigma_{n,n}^{R,<} + G^<_{n,n}\,
\Sigma^{R,a}_{n,n},
\end{equation}
and
\begin{equation}
U_{n,n+1} \, G^<_{n+1,n} = \Sigma_{n,n}^{R,r}\, G^<_{n,n} +
\Sigma^{R,<}_{n,n}\, G^a_{n,n},
\end{equation}
it is straightforward to show that $I_{(n-1)\rightarrow n} =
I_{n\rightarrow(n+1)}$, which guarantees current conservation.
%
%
It is also possible to show that the total current leaving any site is
zero.

\section{Dephasing}
\label{sec:dephasing}

The model Hamiltonian \eqref{eq:tight-bindingH} describes electrons
within the single-particle approximation, disregarding the effects of
ele\-ctron-phonon and electron-electron interactions. Let us call as
``environment" all the degrees of freedom that couple to electrons in
the real physical device. Even at very low temperatures, due to the
interaction with the environment, the quantum interference between
different electronic paths typically fades away at lengths larger than
the scale $\ell_\varphi$. The latter is called coherence length or
dephasing length. Currently, $\ell_\varphi$ in graphene experiments
can be as high as few microns at low temperatures, decreasing with
increasing temperature.
 
This Section shows how to incorporate dephasing into the RGF
method. We follow the phenomenological approach pioneered by
B\"uttiker \cite{Buttiker88} and D'Amato and Pastawski \cite{DAmato90}
and introduce dephasing in the calculations by adding a set of voltage
probes to the system. These voltage probes act on selected system
sites and have their individual chemical potentials adjusted as not
drain or inject any net current. The voltage probes give rise to
dephasing because the drained electrons are not phase coherent with
the ones injected back. The dephasing length $\ell_\varphi$ is related
to the number of voltage probes and the strength of their coupling to
the system. We focus on the linear transport regime and assume that
all scattering within the voltage probes, albeit incoherent, is
elastic. Therefore we neglect any ``vertical flow'', as defined by
Datta \cite{Datta96}.

The basic linear response equations are
\begin{eqnarray}
I_L & = & {\cal Q}_{LL}\, \mu_L - {\cal Q}_{LR}\, \mu_R -
\sum_{i=1}^{N_\varphi} {\cal Q}_{Li}\, \mu_i, \\ \label{eq:IR} I_R & =
& {\cal Q}_{RR}\, \mu_R - {\cal Q}_{RL}\, \mu_L -
\sum_{i=1}^{N_\varphi} {\cal Q}_{Ri}\, \mu_i, \\ I_i & = & {\cal
  Q}_{ii}^\varphi\, \mu_i - \sum_{i^\prime=1 (i^\prime\neq
  i)}^{N_\varphi} {\cal Q}_{ii^\prime}^\varphi\, \mu_{i^\prime} -
    {\cal Q}_{iR}\, \mu_R - {\cal Q}_{iL}\, \mu_L,
\end{eqnarray}
where $\mu_{R(L)}$ denotes the chemical potential in the right (left)
lead and $\mu_i$ is the chemical potential of the $i$th voltage probe,
$i=1,\ldots N_\varphi$, where $N_\varphi$ is the total number of
voltage probes. The linear transport coefficients ${\cal Q}$ need to
be determined (see below). The left, right, and probe currents $I_L$,
$I_R$, and $I_i$, respectively, are defined as positive when they flow
into the system. Since there is no net probe current, we set $I_i = 0$
for all probes. Therefore,
\begin{equation}
\label{eq:Ii}
{\cal Q}_{ii}^\varphi\, \mu_i = \sum_{i^\prime=1 (i^\prime \neq
  i)}^{N_\varphi} {\cal Q}_{ii^\prime}^\varphi \, \mu_{i^\prime} +
{\cal Q}_{iL}\, \mu_L + {\cal Q}_{iR}\, \mu_R.
\end{equation}
In addition, notice that when all the chemical potentials are
identical, all currents should vanish. Thus,
\begin{eqnarray}
\sum_{i=1}^{N_\varphi} {\cal Q}_{Li} + {\cal Q}_{LR} - {\cal Q}_{LL} &
= & 0, \\ \label{eq:GRi} \sum_{i=1}^{N_\varphi} {\cal Q}_{Ri} + {\cal
  Q}_{RL} - {\cal Q}_{RR} & = & 0, \\ \label{eq:Gphi0} {\cal
  Q}_{ii}^\varphi - \sum_{i^\prime=1(i^\prime\neq i)}^{N_\varphi}
{\cal Q}_{ii^\prime}^\varphi - {\cal Q}_{iL} - {\cal Q}_{iR} & = & 0.
\end{eqnarray}
Using Eqs. (\ref{eq:Ii}) and (\ref{eq:Gphi0}), we can then write
\begin{equation}
\sum_{i^\prime=1}^{N_\varphi} {\cal W}_{ii^\prime}^\varphi \left(
\mu_{i^\prime} - \mu_R \right) = {\cal Q}_{iL}\, \left( \mu_L - \mu_R
\right),
\end{equation}
where ${\cal W}_{ii} = {\cal Q}_{ii}^\varphi$ and ${\cal
  W}_{ii^\prime} = - {\cal Q}_{ii^\prime}^\varphi$ if $i\neq
i^\prime$. Whenever the matrix ${\cal W}$ is invertible, we obtain
\begin{equation}
\label{eq:mui}
\mu_i = \mu_R + \sum_{i^\prime=1}^{N_\varphi} \left( {\cal W}^{-1}
\right)_{ii^\prime}\, {\cal Q}_{i^\prime L}\, \left( \mu_L - \mu_R
\right).
\end{equation}
Substituting Eqs. (\ref{eq:GRi}) and (\ref{eq:mui}) into
Eq. (\ref{eq:IR}), we have
\begin{equation}
I_R = \bar{\cal Q}_{RL} \, \left( \mu_R - \mu_L \right),
\end{equation}
where
\begin{equation}
\label{eq:QRLrenorm}
\bar{\cal Q}_{RL} = {\cal Q}_{RL} + \sum_{i,i^\prime=1}^{N_\varphi}
    {\cal Q}_{Ri}\, \left( {\cal W}^{-1} \right)_{ii^\prime} {\cal
      Q}_{i^\prime L}.
\end{equation}
Notice that we have expressed the right-lead current (which is equal
to minus the left-lead current) in terms of the difference between the
chemical potential in the leads.

The linear transport coefficients entering in Eq.~\eqref{eq:QRLrenorm}
can be obtained from the conductance matrix of the system:
\begin{eqnarray}
\label{eq:QRL}
|e|\, {\cal Q}_{RL} & = & {\cal G}_{RL}, \\ 
\label{eq:QRi} 
|e|\, {\cal Q}_{Ri} & = & {\cal G}_{Ri}, \\ 
\label{eq:QiL} 
|e|\, {\cal Q}_{iL} & = & {\cal G}_{iL}, \\ 
\label{eq:Qii} 
|e|\, {\cal Q}_{ii^\prime}^\varphi & = & {\cal G}_{ii^\prime}, \quad i\neq
i^\prime \\ 
|e|\, {\cal Q}_{ii}^\varphi & = & \sum_{i^\prime=1 (i^\prime \neq i)}^{N_\varphi} 
{\cal G}_{ii^\prime} + {\cal G}_{iL} + {\cal G}_{iR}.
\end{eqnarray}
(This makes $\sum_{i=1}^{N_\varphi} {\cal W}_{ii^\prime} \neq 0$, thus
${\cal W}$ is in principle invertible.) The coefficients ${\cal W}$
are positive. This is consistent with the assumption that currents run
from higher to lower chemical potential. The conductances are
calculated using the Caroli formulas
\begin{eqnarray}
\label{eq:Gcross}
{\cal G}_{RL} & = & \frac{2e^2}{h} \mbox{Tr} \left[ \Gamma_R\,
  G_{N+1,0}^r\, \Gamma_L\, G_{0,N+1}^a \right], \\ {\cal G}_{Ri} & = &
\frac{2e^2}{h} \mbox{Tr} \left[ \Gamma_R\, G_{N+1,n_i}^r\, \Gamma_i\,
  G_{n_i,N+1}^a \right], \\ {\cal G}_{iL} & = & \frac{2e^2}{h}
\mbox{Tr} \left[ \Gamma_i\, G_{n_i,0}^r\, \Gamma_L\, G_{0,n_i}^a
  \right], \\ \label{eq:Gii} {\cal G}_{ii^\prime} & = & \frac{2e^2}{h}
\mbox{Tr} \left[ \Gamma_i\, G_{n_i,n_{i^\prime}}^r\,
  \Gamma_{i^\prime}\, G_{n_{i^\prime},n_i}^a \right].
\end{eqnarray}
Here, we use the pair $(n_i,j_i)$ to denote the site coordinates of
the $i$th voltage probe. The level width probe matrix $\Gamma_i$ can
be obtained much in the same way as $\Gamma_R$ and $\Gamma_L$, namely,
from the probe surface Green's function (see Sec. \ref{sec:mesoscopic}).
Notice that in the absence of magnetic fields, all cross-conductance
matrices are symmetric. In addition, since the coupling matrices
$\Gamma_R$, $\Gamma_L$, and $\Gamma_i$ are all Hermitian and positive,
one can easily show that these conductances are all real and positive.

In practice, we will assume that the set of lattice points attached to
voltage probes is sparse, so that $N_\varphi \ll NM$ and the
computation cost of evaluating the cross conductances is not too
high. The number of voltage probes $N_\varphi$ and the magnitude of
the $\Gamma_i$'s (see Sec.~\ref{sec:selenergyprobe}) determine
$\ell_\varphi$.

\subsection{Self consistency}
\label{sec:selfconst}

The Green's function entering in Eqs. (\ref{eq:Gcross}) to
Eq. (\ref{eq:Gii}) are the {\it exact} ones. Therefore, they have to
take into account the coupling to the voltage probes and,
consequently, should depend on the chemical potentials
$\{\mu_i\}$. These, however, depend on the left and right chemical
potentials {\it and} on the cross-conductance matrices ${\cal Q}$,
which are given by Eqs. (\ref{eq:QRi}) to (\ref{eq:Qii}). Thus, one
can see that this calculation needs to be implemented
self-consistently. There is one trivial case though, namely, when the
bias across the system is zero and $\mu_R = \mu_L$ (equilibrium
condition). In this case, Eq. (\ref{eq:mui}) shows that all
$\mu_i=\mu_R$ and the self-consistency can be trivially
satisfied. Nevertheless, one still needs to include the self-energy of
the voltage probes in the local Green's function of each slice during
the recursive calculation: $h_n \rightarrow h_n + \Sigma_i$, when
$i\in n$.

\subsection{Voltage probe self energy}
\label{sec:selenergyprobe}

The voltage probe partial width $\Gamma_i$ is related to the voltage
probe self energy $\Sigma_i$ in the standard way: $\Gamma_i = i \left(
\Sigma_i^r - \Sigma_i^a \right)$. When each voltage probe is attached
a single site, $\Gamma_i$ and $\Sigma_i^r$ are just complex numbers
and the insertion of the probe self energy into the calculation is
substantially simplified. For a one-dimensional semi-infinite chain
with hopping matrix element $t$ coupled to a system site through a
hopping amplitude $t_\varphi$, the retarded Green's function reads
(see Sec. \ref{sec:analytic-squarelattice})
\begin{equation}
g_i^r(E) = \frac{E-\mu_i}{2t^2} \left[ 1 - \sqrt{1 -
    \frac{4t^2}{(E-\mu_i)^2}} \right ]
\end{equation}
Thus, the retarded surface self energy of the $i$th probe is equal to
\begin{equation}
\Sigma_i^r(E) = t_\varphi^2\, g_i^r(E) = \left( \frac{t_\varphi}{t}
\right)^2\, \frac{E-\mu_i}{2} \left[ 1 - \sqrt{1 -
    \frac{4t^2}{(E-\mu_i)^2}} \right],
\end{equation}
leading to
\begin{equation}
\Gamma_i(E) = \left\{ \begin{array}{lr} 0, & |E-\mu_i| > 2t, \\ \left(
  \frac{t_\varphi}{t} \right)^2\, \sqrt{4t^2 - (E-\mu_i)^2}, &
  |E-\mu_i| < 2t. \end{array} \right.
\end{equation}

The retarded Green's function of a slice which is isolated from other
slices should include the self energy of any attached voltage probe:
\begin{equation}
\left[ g_n^{-1} \right](j,j^\prime) = \left\{ \begin{array}{lr} E -
  h_n(j,j) - \Sigma_i^r(E), & j=j^\prime, \\ E - h_n(j,j^\prime) +
  i0^+, & j \neq j^\prime, \end{array} \right.
\end{equation}
when $n_i = n$ and $j = j_i$.

\subsection{Recursion for dephasing Green's functions}
\label{sec:Gdeph}

In order to evaluate the cross-conductance matrices that enter in
Eq. (\ref{eq:QRL}), it is necessary to evaluate the exact Green's
function between two slices carrying voltage probes, say, $n_1$ and
$n_2$; for instance, see Eq. (\ref{eq:Gii}). This is a computationally
intensive task and can only be carried out in a relatively efficient
way if {\it all} left-to-right or right-to-left local Green's
functions are stored during sweeps {\it and} the exact local Green's
functions at either $n_1$ or $n_2$ have already been calculated and
stored.

Suppose that we want to evaluate the retarded Green's function
$G_{n_1,n_2}$, with $n_1 < n_2$. Here are the steps:

\begin{enumerate}
\item Starting with $G_{n_1,n_1}^L$ and $G_{n_1+1,n_1+1}^L$ obtained
  during the left-to-right sweep, evaluate
\begin{equation}
G_{n_1,n_1+1}^L = G_{n_1,n_1}^L\, U_{n_1,n_1+1}\, G_{n_1+1,n_1+1}^L.
\end{equation}
\item Next, use the resulting Green's function and the stored
  $G_{n_1+2,n_1+2}^L$ to evaluate
\begin{equation}
G_{n_1,n_1+2}^L = G_{n_1,n_1+1}^L\, U_{n_1+1,n_1+2}\,
G_{n_1+2,n_1+2}^L.
\end{equation}
\item Repeat this procedure until $G_{n_1,n_2-1}^L$ is obtained.
\item From the previously calculated $G_{n_2,n_2}$, evaluate
\begin{equation}
G_{n_1,n_2} = G_{n_1,n_2-1}^L\, U_{n_2-1,n_2}\, G_{n_2,n_2}.
\end{equation}
\end{enumerate}

Likewise, for $n_1>n_2$, we follow these steps:
\begin{enumerate}
\item Starting with $G_{n_1,n_1}^R$ and $G_{n_1-1,n_1-1}^R$ obtained
  during the right-to-left sweep, evaluate
\begin{equation}
G_{n_1,n_1-1}^R = G_{n_1,n_1}^R\, U_{n_1,n_1-1}\, G_{n_1-1,n_1-1}^R.
\end{equation}
\item Next, use the resulting Green's function and the stored
  $G_{n_1-2,n_1-22}^L$ to evaluate
\begin{equation}
G_{n_1,n_1-2}^R = G_{n_1,n_1-1}^R\, U_{n_1-1,n_1-2}\,
G_{n_1-2,n_1-2}^R.
\end{equation}
\item Repeat this procedure until $G_{n_1,n_2+1}^R$ is obtained.
\item From the previously calculated $G_{n_2,n_2}$, evaluate
\begin{equation}
G_{n_1,n_2} = G_{n_1,n_2+1}^R\, U_{n_2+1,n_2}\, G_{n_2,n_2}.
\end{equation}
\end{enumerate}

Since no inversions are required in these steps, the calculation
scales as $O(|n_1-n_2| M)$.

\section{Disorder}
\label{sec:disorder}

Disorder is ubiquitous in graphene samples, even in those synthesized
with state-of-the-art technologies. Depending on the synthesis method,
charge density inhomoge\-nei\-ties \cite{Martin08} or substrate
irregularities \cite{Zhang09}, intrinsic and extrinsic ripples
\cite{Ishigami07,Meyer07}, strain fields \cite{Levy10}, surface
molecular adsorption \cite{Chen08}, vacancies \cite{Chen09}, and
irregular edges \cite{Han10}, are unavoidable. The effects of these
different kinds of disorder on transport in graphene have been
addressed by several reviews
\cite{CastroNeto09,Mucciolo10,Peres10,DasSarma11} without exhausting
the subject.

The RGF method is flexible enough to address any of the
above-mentioned types of disorder by a suitable choice of the hopping
$t_{ij}$ and the local potential $V_i$ in the Hamiltonian
\eqref{eq:tight-bindingH}. Since it is based on an atomistic basis,
the RGF method is ideal for studying numerically short-range disorder
effects, whose typical range is of the order of the lattice
spacing. Although not optimized for that purpose, the method can also
be used to address long-range disorder
\cite{Rycerz07,Lewenkopf08,Adam09,Klos10}.

In this Section we present some common disorder models for graphene
and discuss their implementation within the RGF method.

\subsection{Diagonal (scalar) disorder}

Charge density and substrate inhomogeneities can be modeled by adding
a local disordered potential $U({\bf r}_i)$ to the lattice sites in
the sample region. One of the simplest models for $U({\bf r}_i)$ is
constructed as follows: We take ${\cal N}_{\rm imp}$ random lattice
sites $\{{\bf R}_k\}$ uniformly distributed as centers of Gaussian
scatterers with a random amplitude $U_k$ taken from a uniform
distribution over the interval $[-\delta V,\delta V]$. This results in
\begin{equation}
U({\bf r}_{n,j}) = \sum_{k=1}^{{\cal N}_{\rm imp}} U_k\, e^{-|{\bf
r}_{n,j} - {\bf R}_k |^2/2\xi^2},
\end{equation}
where $\xi$ is the range of the potential \cite{Shon98,Rycerz07}. The
concentration of scatterers is $n_{\rm imp} = {\cal N}_{\rm imp}/{\cal
  A}$, where ${\cal A}$ denotes the total area of the sample.

In the limit of a low concentration of scatterers, $n_{\rm imp}^{-1/2}
\gg \xi$, the magnitude of the disorder fluctuations is characterized
by the dimensionless parameter $K_0$, which is defined from the
impurity potential correlation function,
\begin{equation}
\label{eq:disorder}
\langle U({\bf r}_{n,j})\, U({\bf r}_{n^\prime,j^\prime}) \rangle =
\frac{K_0 (\hbar v)^2}{2\pi \xi^2}\, e^{-|{\bf r}_{n,j} - {\bf
r}_{n^\prime,j^\prime}|^2/4\xi^2},
\end{equation}
where $v = \sqrt{3}\, a_0\, t/2\hbar$ is the Fermi velocity and
$\langle \cdots \rangle$ stands for the average over disorder
realizations. It is easy to see that $\langle U({\bf
  r}_{n,j})\rangle=0$. We note that $K_0$ contains information not
only about the relative magnitude of the potential fluctuations,
$\delta V/t$, but also about the scatterers' range and concentration:
A simple calculation yields \cite{Rycerz07}
\begin{equation}
\label{eq:K0}
K_0 \approx \frac{16\pi^2}{9} n_{\rm imp} \left( \frac{\delta V}{t}
\right)^2 \left( \frac{\xi^4}{a_0^2} \right).
\end{equation}
If we now recall that there are two inequivalent atoms per hexagon and
$A_{\rm hex} = \sqrt{3}\, a_0^2/2$, we find that
\begin{equation}
K_0 \approx \frac{64\pi^2}{9\sqrt{3}} \frac{{\cal N}_{\rm imp}}{{\cal
N}} \left( \frac{\delta V}{t} \right)^2 \left( \frac{\xi}{a_0}
\right)^4,
\end{equation}
where the numerical prefactor is approximately 40.5 and ${\cal N}$ is
the total number of lattice sites.

In the continuum limit and using Eq. (\ref{eq:disorder}) together with
the Born approximation (BA), one finds that the transport mean free
path away from the Dirac point is given by 
\begin{equation}
\ell_{\rm tr}^{\rm BA} = \frac{2}{\pi} \frac{\lambda_F}{K_0},
\end{equation}
where $\lambda_F$ is the Fermi wavelength in the graphene sheet
\cite{Shon98} ($\lambda_F \ll \ell_{\rm tr}^{\rm BA}$ for the BA
to hold).

\subsection{Off-diagonal disorder -- strain}
\label{sec:strain}

Let us consider the situations where the graphene sheet is subjected
to strain. We now address the case where the strain field modifies the
carbon-carbon bond lengths, postponing to the next subsection the
discussion of bond distortion due to curvatures.

Modifications in the bond lengths lead to a hopping
renormalization. In the Slater-Koster scheme, the carbon-carbon
hopping term can, in principle, be obtained from the dependence of the
$V_{pp\pi}$ on the inter-orbital distance. In practice, one relies on
semi-empirical parameterizations, such as \cite{Pereira09}
\begin{equation}
\label{eq:delta-hopping}
V_{pp\pi}(l) = - t e^{-3.37(l/a - 1)}
\end{equation}
where $l$ is the bond length and $a$ is the inter-atomic distance in
the honeycomb lattice. The decay rate is adjusted to fit the
experimental result $d V_{pp\pi}/dl = -6.4$ eV/\AA. With the help of
Eq.~\eqref{eq:delta-hopping}, local bond length deformations $\delta
l_{i,j} = l_{i,j} - a$ are translated into changes in the hopping
integrals. The latter are easily accounted for by the RGF method.

The macroscopic theory of elasticity can help to translate the strain
field acting on a graphene sheet into modifications of the hopping
integrals. That is because the tensions along the graphene membrane
change very slowly on the microscopic scale. Hence, the changes in the
bond lengths $\delta l_{i,j}$ can be approximated by a smooth function
$\delta l (x,y)$, where $(x,y)$ is the position of the $i$th site in
the honeycomb lattice. In turn, the $\delta l(x,y)$ can be related to
the strain fields by the elastic theory
\cite{Landau}. Reference~\cite{Pereira09}, for instance, writes the
strain tensor ${\bm \varepsilon}$ for the case of uniaxial strain and
shows how to relate ${\bm \varepsilon}$ to the bond length
deformations.

\subsection{Off-diagonal disorder -- ripples}
\label{sec:ripples}

Let us assume that there is a ripple structure in the graphene sheet.
The ripples can be described by a scalar field $h(x,y)$ which
represents the out-of-plane displacement of the carbon atoms at a
given location $(x,y)$. A non-homogeneous $h(x,y)$ ripple-field
modifies the atomic orbital overlaps and, hence, the hopping terms in
the tight-binding model.

Neglecting bond length stretching due to strain, the ripples affect
the nearest neighbor and next-to-nearest neighbor hopping matrix
elements: $t_{ij} = t_{ij}^{(0)} + \delta t_{i,j}$, where
$t_{ij}^{(0)}$ is the hopping between sites $i$ and $j$ in the absence
of ripples and \cite{Kim08}
\begin{equation}
\delta t_{ij} \approx - \frac{1}{2} E_{ij} \left[ (\vec{u}_{ij} \cdot
\nabla) \nabla h \right]^2.
\end{equation}
Here, $\vec{u}_{ij}$ is the unit vector connecting sites $i$ and $j$
and $E_{ij} = t_{ij}^{(0)}/3 + V_{pp\sigma;ij}/2$, where
$V_{pp\sigma;ij}$ describes the overlap of the $\sigma$-orbitals (the
effect of the $\sigma$ orbitals is only negligible in the absence of
bending). The scales are the following:
\begin{equation}
t_{ij}^{(0)} = \left\{ \begin{array}{ll} -2.7\, {\rm eV}, \hskip0.5cm
  & \mbox{for \, n.n.}  \\ -0.1\ {\rm eV}, & \mbox{for \,
    n.n.n.}, \end{array} \right.
\end{equation}
and
\begin{equation}
V_{pp\sigma;ij} = \left\{ \begin{array}{ll} 5.8\, {\rm eV},
  \hskip0.5cm & \mbox{for \, n.n.}  \\ 1.4\ {\rm eV}, & \mbox{for \,
    n.n.n.}. \end{array} \right.
\end{equation}

The free energy associated to the ripple field $h({\bf r})$ for
graphene on a substrate is given by the Gaussian (elastic) form
\cite{Kim08}
\begin{equation}
\label{eq:freeenergy}
F = \frac{1}{2} \int d^2r\, \left[ \kappa (\nabla^2 h)^2 + \gamma
  (\nabla h)^2 + v(h-s)^2 \right],
\end{equation}
where $\kappa$ is the bending rigidity, $\gamma$ is the interfacial
stiffness, and $v$ is a coupling constant for the pinning of the
graphene sheet by the background roughness $s(x,y)$. Typically,
$\kappa \approx 1$ eV. The other two parameters, $\gamma$, and $v$,
will depend on the substrate.

In order to generate the appropriate random $h({\bf r})$, we will
assume that the temperature is low and thermal fluctuations can be
neglected (this hypothesis could in principle be relaxed). In this
case, the fluctuations in $h({\bf r})$ follow those of the substrate:
Minimizing the free energy in Eq. (\ref{eq:freeenergy}), we obtain
\begin{equation}
\left( \kappa \nabla^4 - \gamma \nabla^2 + v \right) h(x,y) = v\,
s(x,y).
\end{equation}
Using the Fourier decompositions $h(\vec{r}) = \sum_{\vec{q}}
\tilde{h}(\vec{q})\, e^{i \vec{q} \cdot \vec{r}}$ and $s(\vec{r}) =
\sum_{\vec{q}} \tilde{s}(\vec{q})\, e^{i \vec{q} \cdot \vec{r}}$ we
obtain
\begin{equation}
\left( \kappa q^4 + \gamma q^2 + v \right) \tilde{h}(\vec{q}) = v\,
\tilde{s}(\vec{q}).
\end{equation}
As a result,
\begin{equation}
\left\langle \tilde{h}(\vec{q}_1)\, \tilde{h}(\vec{q}_2) \right\rangle
= \frac{v^2\, \left\langle \tilde{s}(\vec{q}_1)\, \tilde{s}(\vec{q}_2)
\right\rangle} {(\kappa q_1^4 + \gamma q_1^2 + v) (\kappa q_2^4 +
\gamma q_2^2 + v)}.
\end{equation}
If the background has white-noise fluctuations, $\langle
\tilde{s}(\vec{q}_1)\, \tilde{s}(\vec{q}_2) = s_0^2\, \delta^{(2)}
(\vec{q}_1 + \vec{q}_2)$, we get
\begin{equation}
\left\langle \tilde{h}(\vec{q}_1)\, \tilde{h}(\vec{q}_2) \right\rangle
= \frac{v^2\, s_0^2\, \delta^{(2)}\left( \vec{q}_1 + \vec{q}_2
\right)} {(\kappa q_1^4 + \gamma q_1^2 + v)^2}.
\end{equation}
Since
\begin{equation}
\left\langle h(\vec{r})\, h(0) \right\rangle =
\sum_{\vec{q}_1,\vec{q}_2} \left\langle \tilde{h}(\vec{q}_1)\,
\tilde{h}(\vec{q}_2) \right\rangle\, e^{i \vec{q}_1 \cdot \vec{r}},
\end{equation}
we finally arrive at
\begin{equation}
\label{eq:correlation}
\left\langle h(\vec{r})\, h(0) \right\rangle = \sum_{\vec{q}}
\frac{v^2\, s_0^2} {(\kappa q^4 + \gamma q^2 + v)^2} e^{i \vec{q}
\cdot \vec{r}}.
\end{equation}
We now need to find a way to generate membrane profiles $h(x,y)$ that
satisfy this correlation function. The solution is simple: Let us
introduce
\begin{equation}
\label{eq:hc}
h_c(\vec{r}) = \sum_{\vec{q}} \frac{v\, s_0} {\kappa q^4 + \gamma q^2
+ v} e^{i (\vec{q} \cdot \vec{r} + \phi_{\vec{q}})},
\end{equation}
where the phases $\left\{\phi_{\vec{q}} \right\}$ are uniformly
distributed in the interval [0:2$\pi$) and uncorrelated, except that
  $\phi_{\vec{q}} = - \phi_{-\vec{q}}$ in order to define a real
  $h_c$. It then follows that
\begin{align}
\label{eq:hcdecomp}
\left\langle h_c(\vec{r})\, h_c(0) \right\rangle & = 
\sum_{\vec{q}_1,\vec{q}_2} \frac{v^2\, s_0^2} {(\kappa q_1^4 + \gamma
q_1^2 + v) (\kappa q_2^4 + \gamma q_2^2 + v)} \nonumber\\
& \hskip1.0cm \times e^{i \vec{q}_1 \cdot
\vec{r}} \left\langle e^{i (\phi_{\vec{q}_1} + \phi_{\vec{q}_2})}
\right\rangle \nonumber \\ & = \sum_{\vec{q}} \frac{v^2\, s_0^2} {(
\kappa q^4 + \gamma q^2 + v)^2} e^{i \vec{q} \cdot \vec{r}},
\end{align}
which is exactly equal to Eq. (\ref{eq:correlation}).

Equation (\ref{eq:hc}) needs to be adapted to a strip geometry. First,
it is clear that, in rectangular coordinates,
\begin{equation}
h_c(x,y) = \sum_{q_x} \sum_{q_y \geq 0} \frac{2v\, s_0} {\kappa q^4 +
\gamma q^2 + v} \cos \left( x q_x + y q_y + \phi_{q_x,q_y} \right).
\end{equation}
Moreover, we can assume $q_x = 2\pi n_x/N_x$ and $q_y = 2\pi n_y/N_y$,
with $n_x = -N_x/2, \ldots, (N_x-1)/2$ and $n_y = -N_y/2, \ldots,
(N_y-1)/2$, where $N_x \times N_y$ is the number of grid points in
real space. One could write the real space grid using the primitive
lattice vectors of the hexagonal (actually triangular) underlying
system. However, since the strip has a rectangular geometry and the
field $h(x,y)$ is defined over a coarse grained lattice (hydrodynamic
continuum limit) which does not need to reflect the underlying atomic
structure. Thus, a rectangular mesh suffices and we can rewrite
Eq. (\ref{eq:hcdecomp}) as
\begin{align}
h_c(x,y) = \sum_{n_x=-\frac{N_x}{2}}^{\frac{N_x-1}{2}} & \sum_{n_y =
  0}^{\frac{N_y-1}{2}} \, \frac{2v\, s_0} {\kappa q^4 + \gamma q^2 +
  v}\nonumber\\ & \hskip-0.3cm \times \cos \left[ 2\pi \left( x
  \frac{n_x}{N_x} + y \frac{n_y}{N_y} \right) + \phi_{n_x,n_y}
  \right],
\end{align}
with $q^2 = (2\pi)^2 [ (n_x/N_x)^2 + (n_y/N_y)^2]$, where we have
implicitly assumed $N_y$ to be odd.

\subsection{Vector potential (off-diagonal) disorder}
\label{sec:offdiagonal}

The off-diagonal disorder discussed in the previous Sections can be
cast in terms of a random vector potential, as nicely reviewed in
Ref.~\cite{Vozmediano10}. Here we briefly present the mapping of the
tight-binding hopping disorder $\delta t_{ij}$ into a random vector
potential ${\bf A}(x,y)$. We then show how to implement this kind of
disorder in the tight-binding model.

The continuum limit of the honeycomb lattice tight-bin\-ding model
near the neutrality points translates into a Dirac equation. Using the
Bloch states of one (A or B) of the triangular sublattices (i.e., the
underlying Bravais lattice) that constitute honeycomb lattice, one
obtains the Hamiltonian
\begin{equation}
\label{eq:singlecone}
\hat{H}_{\rm AB} = t \left( \begin{array}{cc} 0 & 1 +
e^{i{\bf k} \cdot {\bf a}_1} + e^{i{\bf k} \cdot {\bf a}_2}
\\ 1 + e^{-i{\bf k} \cdot {\bf a}_1} + e^{-i{\bf k} \cdot {\bf
a}_2} & 0
\end{array} \right),
\end{equation}
where ${\bf a}_1 = (a_0/2,a_0\sqrt{3}/2)$ and ${\bf a}_2 =
(-a_0/2,a_0\sqrt{3}/2)$ are the primitive (Bravais) lattice
vectors. Here we use the lattice vector conventions of
Ref.~\cite{CastroNeto09}. The Hamiltonian $H$ acts on a spinor whose
components are the envelop wave function amplitudes at the sublattices
A and B. The corresponding low-energy dispersion relation shows two
inequivalent cones (valleys) centered at the $k$-space points
\begin{equation}
{\bf K} = \left(-\frac{4\pi}{3a_0},0 \right) \qquad {\rm and} \qquad
{\bf K}^\prime = \left(\frac{4\pi}{3a_0},0 \right).
\end{equation}
Expanding ${\bf k}$ around these points, one obtains the Hamiltonians
\begin{align}
\hat{H}_K = & v\hbar \left( \begin{array}{cc} 0 & k_x - ik_y \\ k_x
+ ik_y & 0 \end{array} \right) = v\hbar\, (k_x \hat{\sigma}_x + k_y
\hat{\sigma}_y) \\ \hat{H}_{K^\prime} = & v\hbar \left(
\begin{array}{cc} 0 & -k_x - ik_y \\ -k_x + ik_y & 0
\end{array} \right)  = v\hbar\, (-k_x \hat{\sigma}_x + k_y \hat{\sigma}_y),
\end{align}
where $k_x$ and $k_y$ are measured from the cone vertices. The
real-space, continuum version of the Hamiltonians can be obtained by
replacing $k_x$ with $-i\partial_x$ and $k_y$ with $-i\partial_y$. As
a result,
\begin{eqnarray}
\hat{H}_K & = & v\hbar \left( \begin{array}{cc} 0 & -i\partial_x -
  \partial_y \\ -i\partial_x + \partial_y & 0 \end{array} \right) \\
  \hat{H}_{K^\prime} & = & v\hbar \left( \begin{array}{cc} 0 &
  i\partial_x - \partial_y \\ i\partial_x +\partial_y & 0
\end{array} \right).
\end{eqnarray}
When a vector potential is present, the substitution is instead
$k_{x,y} \longrightarrow -i\partial_{x,y} + \frac{e}{\hbar c}
A_{x,y}$ (with $e>0$) and the Hamiltonians become
\begin{align}
\hat{H}_K \!= & v\hbar \left( \begin{array}{cc} 0 & \hskip-0.5cm
  -i\partial_x -\partial_y +\frac{e}{\hbar c} (A_x -iA_y)
  \\ -i\partial_x + \partial_y + \frac{e}{\hbar c} (A_x +iA_y)
  & \hskip-0.5cm 0 \end{array} \right) \\ \hat{H}_{K^\prime} \! = &
v\hbar \left( \begin{array}{cc} 0 & \hskip-0.2cm i\partial_x -
  \partial_y - \frac{e}{\hbar c} (A_x + iA_y) \\ i\partial_x +
  \partial_y - \frac{e}{\hbar c} (A_x - iA_y) & \hskip-0.2cm 0
\end{array} \right).
\end{align}

We now show how a local (long-ranged) distortion in the lattice gives
raise to $H_{K(K')}$ as above. Let us assume that the three
nearest-neighbor hopping amplitudes of any given site are not equal:
Calling them $t_0$, $t_1$, and $t_2$, we have to rewrite
Eq. (\ref{eq:singlecone}) in the form
\begin{equation}
\hat{H}_{\rm AB} = \left( \begin{array}{cc} 0 & t_0 + t_1
e^{i{\bf k} \cdot {\bf a}_1} + t_2 e^{i{\bf k} \cdot {\bf a}_2} \\ t_0
+ t_1 e^{-i{\bf k} \cdot {\bf a}_1} + t_2 e^{-i{\bf k} \cdot {\bf
a}_2} & 0
\end{array} \right).
\end{equation}
Expanding around the same ${\bf K}$ and ${\bf K}^\prime$ points, and
assuming $|t_1-t_2| \ll |t_1+t_2|$, we find
\begin{align}
\hat{H}_K =& \; v\hbar \left( \begin{array}{cc} 0 & k_x - ik_y \\ k_x
+ ik_y & 0 \end{array} \right) \nonumber\\ 
& \hskip-0.4cm+ \left( \begin{array}{cc} 0 &  \hskip-0.2cm t_0 -
\frac{t_1+t_2}{2} - \frac{i \sqrt{3} (t_1-t_2)}{2} \\ t_0 -
\frac{t_1+t_2}{2} + \frac{i \sqrt{3} (t_1-t_2)}{2} & \hskip-0.2cm 0 \end{array}
\right)
\end{align}
\begin{align}
\hat{H}_{K^\prime} = & \; v\hbar \left( \begin{array}{cc} 0
& -k_x - ik_y \\ -k_x + ik_y & 0
\end{array} \right)  \nonumber\\ 
& \hskip-0.4cm+ \left( \begin{array}{cc} 0 & \hskip-0.2cm t_0 -
\frac{t_1+t_2}{2} + \frac{i \sqrt{3} (t_1-t_2)}{2} \\ t_0 -
\frac{t_1+t_2}{2} - \frac{i \sqrt{3} (t_1-t_2)}{2} & \hskip-0.2cm 0 \end{array}
\right).
\end{align}
Thus, we can define two vector potentials, one for each cone:
\begin{equation}
\label{eq:vecpotK}
A_x^K = \frac{c}{ve} \left( t_0 - \frac{t_1+t_2}{2} \right), \quad
A_y^K = \frac{c}{ve} \frac{\sqrt{3}}{2} ( t_1-t_2),
\end{equation}
and
\begin{equation}
A_x^{K^\prime} = -\frac{c}{ve} \left( t_0 - \frac{t_1+t_2}{2} \right),
\quad A_y^{K^\prime} = -\frac{c}{ve} \frac{\sqrt{3}}{2} ( t_1-t_2).
\end{equation}
Notice that ${\bf A}^{K} = - {\bf A}^{K^\prime}$, as expected from
time-reversal symmetry considerations.

We can use the vector potential as a gauge field that parameterizes
local fluctuations in the nearest-neighbor hopping matrix
elements. For this purpose, it is useful make a single-cone
approximation and rewrite Eq. (\ref{eq:vecpotK}) in the form
\begin{eqnarray}
\label{eq:offdiagdis0}
t_0(n,j) & = & t + \frac{2ve}{c}\, A_x(n,j), \\ 
\label{eq:offdiagdis1}
t_1(n,j) & = & t + \frac{ve}{c}\, \left[ A_x(n,j) + \frac{1}{\sqrt{3}}
A_y(n,j) \right], \\ 
\label{eq:offdiagdis2}
t_2(n,j) & = & t + \frac{ve}{c}\, \left[ A_x(n,j) - \frac{1}{\sqrt{3}}
  A_y(n,j) \right],
\end{eqnarray}
where $(n,j)$ are the coordinates of sites belonging to one of the
sublattices. Notice that this type of off-diagonal disorder does break
particle-hole symmetry.

Another way to proceed and get the same results is to use a Peierls
substitution in the hopping matrix elements \cite{peierls}, such that
\begin{eqnarray}
e^{i{\bf k} \cdot {\bf a}_1} & \longrightarrow & e^{i\left( {\bf k} +
\frac{e}{\hbar c} {\bf A} \right) \cdot {\bf a}_1} \\
e^{i{\bf k} \cdot {\bf a}_2} & \longrightarrow & e^{i\left( {\bf k} +
\frac{e}{\hbar c} {\bf A} \right) \cdot {\bf a}_2}
\end{eqnarray}
in Eq. (\ref{eq:singlecone}). For instance, expanding on {\bf A} and
around the ${\bf K}$ point, we obtain
\begin{align}
t & \left[ 1 + e^{i\left( {\bf k} + \frac{e}{\hbar c} {\bf A} \right)
\cdot {\bf a}_1}  + e^{i\left( {\bf k} + \frac{e}{\hbar c} {\bf A}
\right) \cdot {\bf a}_2} \right] \approx \nonumber \\
& \hskip2.4cm \hbar v \left(k_x - i k_y
\right) + \frac{ev}{c} \left( A_x - iA_y \right).
\end{align}
Then, setting
\begin{equation}
t_0 - \frac{t_1+t_2}{2} - i\sqrt{3}\, \frac{t_1-t_2}{2} = \frac{ev}{c}
\left( A_x - iA_y \right),
\end{equation}
we arrive at Eqs. (\ref{eq:offdiagdis0}) to (\ref{eq:offdiagdis2}).

This concludes the demonstration that a hopping distortion can be
mapped into a corresponding vector field. From the point of view of a
tight-binding modeling and the RGF method, it seems simpler to model
strain with renormalized hoppings, thus avoiding issues related to
projections onto the ${\bf K}$ and ${\bf K}'$ cones to preserve
time-reversal symmetry.

The random gauge potential model is also interesting for other
reasons. It has a physical realization in rippled graphene subjected
to a strong parallel magnetic field \cite{Lundeberg10} and has been
analytically addressed by several authors, e.g.
Ref. \cite{Ostrovsky06}.

Let us model the random vector potential by assuming that the gauge
field has Gaussian fluctuations, such that
\begin{equation}
\left\langle A_\alpha(n,j)\, A_\beta(n^\prime,j^\prime) \right\rangle
= \lambda\, \delta_{\alpha\beta}\, e^{-|{\bf r}_{n,j} - {\bf
r}_{n^\prime,j^\prime}|^2/2\xi^2},
\end{equation}
where $\lambda$ measures the strength of the fluctuations and $\xi$ is
their correlation length. One way to generate this correlation
function is to define at the nodes of a regular lattice of constant
$a_g$ two sets of uniformly distributed random numbers\footnote{One
  set of random numbers for each $\alpha$-component of the gauge
  field, such that $\langle c_k^\alpha \rangle = 0$ and $\langle
  c_k^\alpha c_{k^\prime}^{\alpha^\prime} \rangle = \delta_{kk^\prime}
  \delta_{\alpha\alpha^\prime}$.} $\{ c_k^\alpha \}_{k=1,\ldots,{\cal
    N}}$ and to define the gauge field through the expression
\begin{equation}
A_\alpha(n,j) = \frac{f}{\cal C} \sum_{k=1}^{\cal N} c_k^\alpha
e^{-|{\bf r}_{n,j} - {\bf R}_k|^2/\xi^2},
\end{equation}
where
\begin{equation}
{\cal C} = \sum_{k=1}^{\cal N} e^{-|{\bf r}_{n,j} - {\bf R}_k|^2/\xi2}
\longrightarrow \pi \left(\frac{\xi}{a_g} \right)^2
\end{equation}
and $\lambda = f^2(a_g/\xi)^2/2\pi$ when ${\cal N} \rightarrow \infty$
maintaining ${\cal N}a_g < \infty$. This construction implicitly
assumes that $a_g \ll \xi$.

There is a useful way to quantify the fluctuations of the vector
potential. Let us denote $\left\langle \Phi^2 \right\rangle$ the rms
value of the magnetic flux piercing a region of area ${\cal A} <
\xi^2$. Then, $\left\langle \Phi^2 \right\rangle \approx {\cal A}^2
\left\langle B^2 \right\rangle$, where $B = \partial_x A_y -
\partial_y A_x$. Following the steps shown in Appendix \ref{sec:flux},
we find that $\left\langle B^2 \right\rangle = f^2
(a_g/\xi)^2/2\pi\xi^2$.

We can now redefine the vector potential to absorb the prefactor
$e/\hbar c$: $\tilde{\bf A} \equiv (e/\hbar c) {\bf A}$. Likewise, in
order to get rid of the prefactor in the expressions used to generate
the vector potential, we introduce $\tilde{f} \equiv (\hbar c/e) f$
and $\tilde{\lambda} \equiv (\hbar c/e)^2 \lambda$. Then, we can write
the following expression for the estimated rms value of the random
magnetic flux in units of the flux quantum ($\Phi_0 = hc/e$):
\begin{equation}
\frac{\delta\varphi}{2\pi} \equiv \frac{\sqrt{\left\langle \Phi^2
\right\rangle}}{\Phi_0} \approx \frac{1}{2\pi}\, \frac{1}{\sqrt{2\pi}}
\left( \frac{{\cal A}}{\xi^2} \right)\, a_g\, \tilde{f},
\end{equation}
which implies $\tilde{\lambda} = (\delta\varphi)^2 \xi^2/{\cal
A}^2$. Thus, the relation between the rms flux phase piercing an
elementary hexagon ${\cal A}_{\rm hex} = \sqrt{3}\, a_0^2/2$ and the
vector potential intensity is
\begin{equation}
\tilde{f}_{\rm hex} = \sqrt{\frac{8\pi}{3}}\, \left( \frac{\xi}{a_0}
\right)^2 \frac{\delta\varphi}{a_g},
\end{equation}
or, equivalently,
\begin{equation}
\frac{\tilde{f}_{\rm hex}}{{\cal C}} = \sqrt{\frac{8}{3\pi}}\, \left(
\frac{a_g}{a_0} \right)^2 \frac{\delta\varphi}{a_g}.
\end{equation}
On the other hand, if we set ${\cal A}_{\rm ripple} = \xi^2$ to denote
the typical area of a ripple, we obtain
\begin{equation}
\tilde{f}_{\rm ripple} = \sqrt{2\pi}\, \frac{\delta\varphi}{a_g},
\qquad \frac{\tilde{f}_{\rm hex}}{{\cal C}} = \sqrt{\frac{2}{\pi}}\,
\left( \frac{a_g}{\xi} \right)^2 \frac{\delta\varphi}{a_g}.
\end{equation}
%

\subsection{Edge disorder}
\label{sec:nanoribbons}

Etching a graphene sheet to produce nanoribbons always leaves behind
some roughness at the edges. When the irregular shape of the
boundaries of the propagating region is very pronounced, it leads to
the formation of ``bottlenecks'' and ``cavities'', which tend to
increase charging effects and lead to Coulomb blockade oscillations of
the conductance \cite{Sols07}. However, even mild amounts of edge
disorder can affect dramatically electronic transport in
nanoribbons. In this case, it has been proposed that for long enough
nanoribbons, Anderson localization (and thus an insulating behavior)
can develop \cite{Verges07,Evaldsson08,Mucciolo09}. Insulating
behavior, albeit of a different nature, is also expected in
``perfect'' nanoribbons due to lattice symmetric breaking caused by
the deformation of the chemical bonds involving carbon atoms at the
edges, as revealed by DFT calculations \cite{Son06}

Edge disorder can be simulated by considering slices with random
numbers sites (see Fig.~\ref{fig:nanoribbon}): For instance, we can
draw the number of sites $M_n$ of the $n$th slice randomly according
to the Gaussian distribution
\begin{equation}
P(M_n) = \frac{1}{\sqrt{2\pi} \delta M} \,
e^{-\left(M_n-\overline{M}\right)^2/2\delta M^2}
\end{equation}
where $\overline{M}$ is the average number of transverse unit cells
in the nanoribbon and $\delta M$ is its standard deviation. Other
distributions can be investigated straightforwardly. The slices are
concatenated such that hopping matrix elements connecting sites which
fall into empty spaces are set to zero (although this can be avoided
when programming the recursive calculation by using nested loops with
variable ranges). Thus, one may think of this procedure as a random
removal of sites at the edges of the nanoribbon. This approach has
been used to study the existence of localized states in graphe\-ne
systems \cite{Verges07}.

Another approach, which tries to mimic the effect of etching, is
explained in \cite{Mucciolo09}. Again, the numbers $\{M_n\}$ are
considered random variables, but their generation follows a different
procedure. One visits sequentially each edge site (at the top and
bottom) and elects to removes it (or not) according to a probability
$p_1$. Certain sites, when removed, require the removal of neighboring
sites as well, as an edge configuration where carbon atoms have a
single bond are not stable (unless both dangling bonds in the carbon
atom are pacified, but this is not likely to occur during
etching). After this first sweep of edge sites, a second sweep
follows, but now sites are removed with a probability $p_2$. One can
continue repeating this procedure, using a different removal
probability at each sweep, until the desired amount of roughness is
obtained.

\section{Some Numerical Results}
\label{sec:numerics}

In this Section we show some representative results obtained with the
RGF method. We begin by addressing the case of ballistic transport in
graphene sheets where analytical results are known and serve to
benchmark the numerical method. Next, we discuss the case of graphene
sheets with long-range disorder, where the RGF method was used to
clarify the controversial issue of the ``universal conductivity
minimum". Finally, we present results for graphene nanoribbons,
showing how the method can be used to calculate LDOS and the local
current density.

\begin{figure}[h]
\centering
\includegraphics[width=0.85\columnwidth]{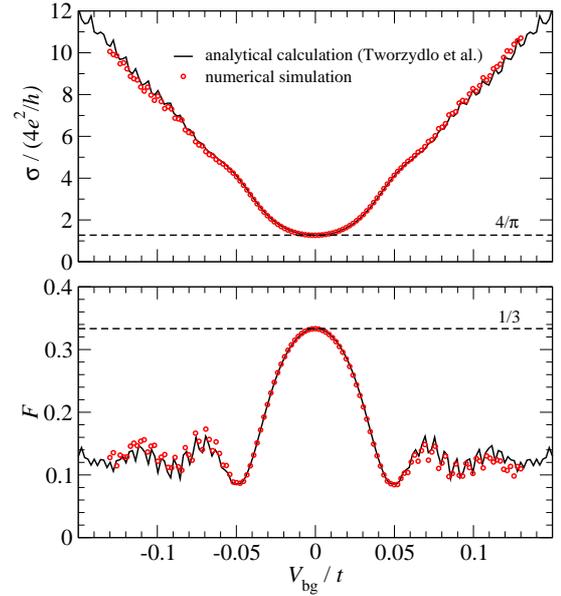}
\caption{Results of linear transport calculations for clean graphene
  sheets: Conductivity (defined as $\sigma=LG/W$) and Fano factor as a
  function of the Fermi energy in the contacts ($V_g = 0$). The band
  in the square lattice leads was offset such that its middle
  coincides with the neutrality point in the graphene sheet in order
  to increase the density of states at the contact and thus mimic a
  metallic lead. Armchair edges, $M=360$ and $N=70$ (aspect ratio $W/L
  = 5.2$).}
\label{fig:ball}
\end{figure}

\subsection{Ballistic transport in clean samples}

Here we present results obtained for the case of ballistic transport
in graphene sheets. We consider mainly the armchair orientation, since
this, in the clean limit, provides a band structure and dispersion
relation very similar to a quasi-one-di\-mensional projection of the
Dirac fermion model and is more suitable for scaling analyzes.

First, in Fig. \ref{fig:ball}, we show results for the clean limit (no
bulk or edge disorder) for a short ribbon, keeping the back gate
voltage fixed to zero (neutrality point) and varying the Fermi energy
in the contacts (zero bias). The numerical data is compared to the
analytical expressions derived by in Ref. \cite{Tworzydlo06}. The
agreement is quite good for large systems and becomes worse when the
system is too small (not shown). In particular, a strong asymmetry and
a lack of well-defined oscillations occurs if the system is not large
enough (not shown).

\begin{figure}[t]
\centering
\includegraphics[width=0.95\columnwidth]{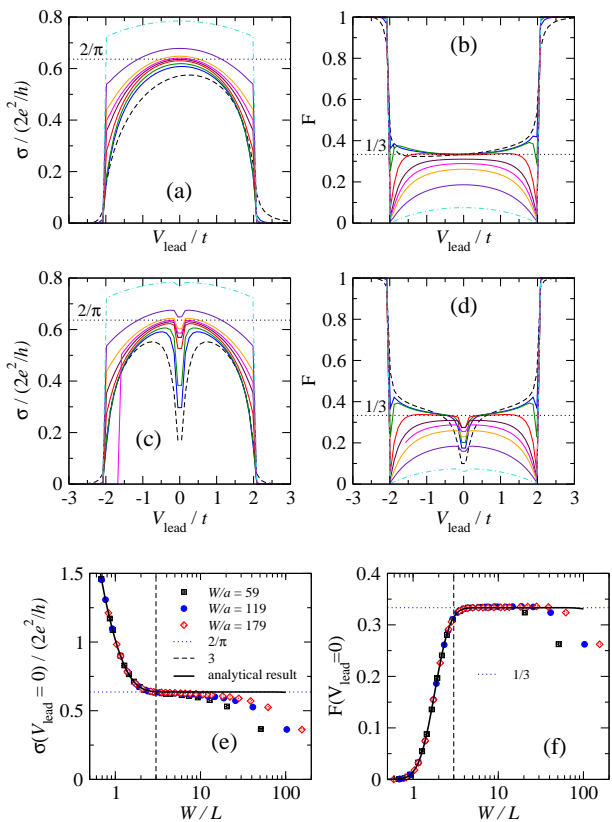}
\caption{Results of numerical simulations of clean ribbons:
  conductivity and Fano factor as a function of the Fermi energy in
  the contacts ($V_g = 0$) for different aspect ratios. (a), (b), (e),
  (f): Square lattice contacts; (c) and (d): Armchair honeycomb
  contacts. For all plots, $M=120$. In plots (a), (b), (c), and (d),
  the value of $N$ are 12 (dashed line), 24, 36, 72, 96, 108, 120,
  148, and 200 (dashed-dotted line). The thick solid line corresponds
  to the analytical result \cite{Tworzydlo06}.}
\label{fig:comparison}
\end{figure}

The ability of the recursive method to get precise results for clean
systems is clear also in Fig. \ref{fig:comparison}, where the
calculations are performed for different aspect ratios. The deviations
from the analytical curve only occur when the system is too short and
evanescent modes dominate transport.

\begin{figure}[h]
\centering
\includegraphics[width=0.90\columnwidth]{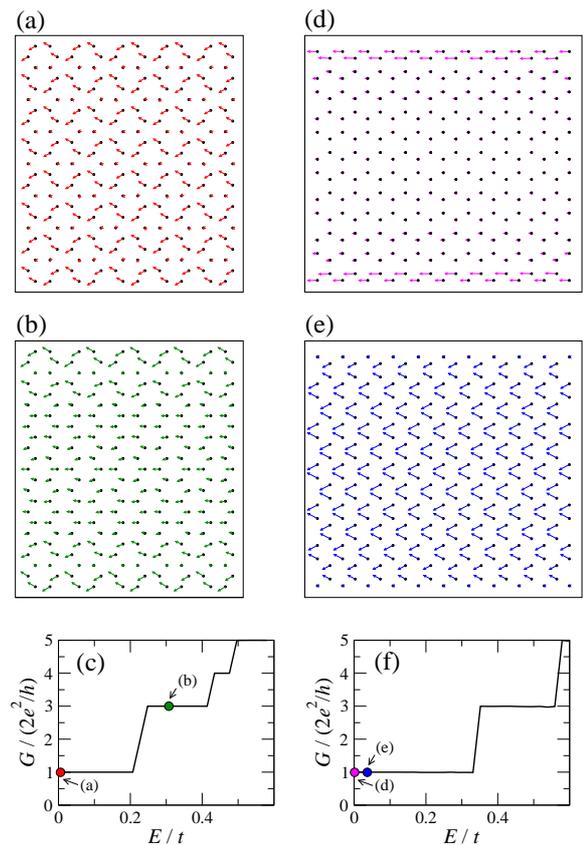}
\caption{Current densities (a,b,d,e) and linear conductance steps
  (c,d) of two small ballistic graphene ribbons. In (a,b,c,d) the
  arrows represent the current densities (in arbitrary units)
  evaluated at different sites using Eq. (\ref{eq:Idensity}). Armchair
  edges, $M=12$, $N=20$: (a), (b), and (c); zigzag edges, $M=12$,
  $N=21$: (d), (e), and (f). Energies: $E=0$ for plots (a) and (d);
  $E=0.3t$ for plot (b); $E=0.01t$ for plot (e).
  amplitude.}
\label{fig:current_dens}
\end{figure}

In Figs. \ref{fig:current_dens} we show the conductance and the
current density in the linear regime for small flakes with armchair
and zigzag edges. Here, the leads are also honeycomb lattices. In this
case, the conductance steps can be easily understood from the energy
dispersion relation of graphene infinite ribbons \cite{Brey06},
namely, the dimensionless conductance $G/(2e^2/h)$ is given by the
number of bands crossing the Fermi energy $E$.

There is no such simple explanation for the current density. Notice
that the notable difference in the current distribution for armchair
and zigzag orientations at $E=0$: While in the latter the current is
primarily carried by edge states, in the former the current is
uniformly distributed across the flake. As one moves a just little bit
away from $E=0$, the current distribution for the zigzag flake changes
drastically, with nearly no current running at the edges. This result
is related to the the fact that for zigzag nanoribbons the $E=0$
states are strongly localized at the edges. As soon as $|E|>0$ both
edge states and edge currents disappear, even in the case of a single
conducting channel. For the armchair orientation, the change in the
current distribution for increasing energies is less drastic.

It should be stressed the edge current densities of zigzag nanoribbons
change both quantitative and qualitative if one switches from
nearest-neighbor \cite{Nikolic07} to next-nearest-neighbor
tight-binding models \cite{Nikolic12}. The issue of which mo\-del is
appropriate is tied to the desire to fit DFT calculations \cite{Son06}
or to explain experimental manipulation and characterization of
nanoribbon edges \cite{Tao11}.

\subsection{Disordered graphene sheets}

The conductivity minimum $\sigma_0$ observed at the charge neutrality
point in graphene monolayers has been a subject of intense debate,
which is reviewed, for instance, in Ref.~\cite{Mucciolo10}. Here, we
show how the long-range Gaussian correlated potential can be used to
investigate the value of $\sigma_0$. \footnote{The discussion and
  results that follow complement the material presented in
  Ref.~\cite{Mucciolo10}.}

In the diffusive regime, in general, the system geometry has 
little influence on the transport properties which allows one to
express the average conductivity as $\sigma= (L/W) \langle G\rangle$,
where $L$ is the system length and $W$ its width. We use the same
setting as in the previous subsection, including now long-range
Gaussian disorder in the device region. To generate the data shown in
Fig. \ref{fig:conductivity_min}, four different aspect ratios were
considered as well as several values of $K_0$ and $\xi /a$. The
average conductivity $\sigma_0$ obtained from $\langle G(V_g=0)\rangle
$ is plotted versus $L$ scaled by $\ell^*$. The parameter $\ell^*$
depends on $K_0$ and $\xi$. We identify $\ell^*$ with the elastic
disorder mean free path $\ell$.

\begin{figure}[h]
\centering
\includegraphics[width=0.95\columnwidth]{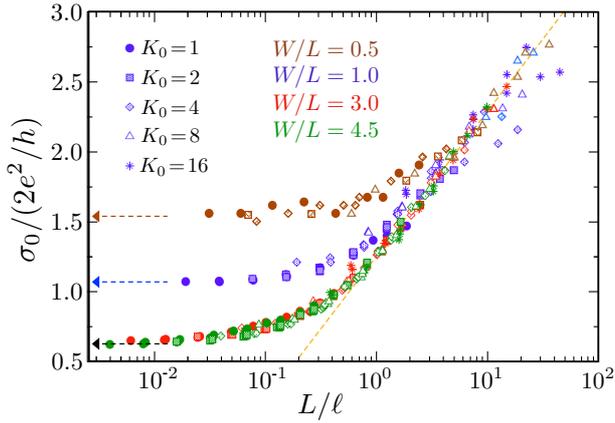}
\caption{Conductivity minimum $\sigma_0$ in unit of $2e^2/h$ as a
  function of the system size $L$ scaled by the electron mean free
  path $\ell$. The results correspond to the average over $10^2 \cdots
  10^4$ disorder realizations for $L$ ranging between 50 and 500
  $a_0$. The colors represent different aspect ratios $W/L$. The
  symbols stand for the values of the dimensionless disorder strength
  $K_0$. The arrows indicate the analytical value of the conductivity
  minimum in the ballistic limit \cite{Tworzydlo06}, which depends on
  $W/L$. The dotted line gives the diffusive $\ln(L/\ell)$ behavior
  \cite{Ostrovsky06}.}
\label{fig:conductivity_min}
\end{figure}

Let us summarize the results shown in
Fig.~\ref{fig:conductivity_min}. Two clear regimes can be
identified. For $L/\ell\ll 1$, the probability of an electron being
scattered by disorder as it traverses the sample is very small. This
corresponds to the ballistic regime, where scattering occurs mainly at
the sample edges and transport properties are dominated by the sample
geometry. Note that when $L/\ell <1$, $\sigma_0$ approaches the
prediction for the pure ballistic case \cite{Tworzydlo06}, indicated
by the arrows in Fig.~\ref{fig:conductivity_min}. In contrast, when
$L/\ell\gg 1$, the system becomes diffusive and geometry affects
transport weakly. Figure \ref{fig:conductivity_min} clearly shows this
crossover. For the diffusive regime, $L/l \gg 1$, the conductivity is
proportional to $\ln(L/\ell)$, in agreement with the non-linear sigma
model prediction \cite{Ostrovsky06}. The mismatch between the
numerical prefactor for the logarithm and the value characteristic of
the symplectic class may be related to the finite contact resistance
\cite{Adam09} present in our simulations.

These simulations suggest an explanation for results obtained in
transport experiments at the charge neutrality point. In the coherent
diffusive regime, the conductivity minimum has significant
sample-to-sample fluctuations and its average shows a weak
(logarithmic) dependence on the mean free path. Typical diffusive
experimental samples have $L/\ell \approx 1 - 10$ and $\sigma_0
\approx 4e^2/h$, similarly to what is shown in
Fig.~\ref{fig:conductivity_min}.

\begin{figure}[t]
\centering
\includegraphics[width=0.8\columnwidth]{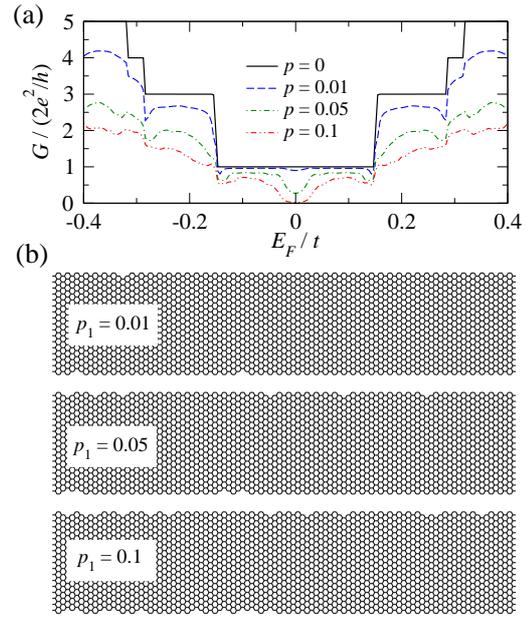}
\caption{(a) Linear conductance of edge disordered nanoribbons with
  armchair edge orientation, $M=18$, $N=200$, averaged over 100
  realizations, as a function of energy. Only one etching sweep is
  used, but results for three different values of the site removal
  probability $p_1$ are shown. (b) Typical realizations used in (a)
  for value of $p_1$ considered.}
\label{fig:nanoribbon}
\end{figure}

\subsection{Nanoribbons}

For nanoribbons, both bulk and edge disorder play a role in electronic
transport. In the absence of band gaps, long-range disorder does not
suppress conductance significantly and a perfect conducting channel
exists near the neutrality point
\cite{Wakabayashi07,Wakabayashi09,Lima12}. The story is quite
different for short-range disorder. Bulk imperfections (lattice
defects, impurities, or adsorbates) and edge imperfections can lead to
strong localization due to backscattering and enhanced destructive
interference \cite{Evaldsson08,Mucciolo09}. To illustrate this point,
Fig. \ref{fig:nanoribbon} shows the rapid smearing of the linear
conductance steps of a nanoribbon when even a small amount of edge
sites are randomly removed (i.e., etched out). This shows how
challenging it is to observe conductance quantization experimentally
in these systems.

\begin{figure}[h]
\centering
\includegraphics[width=0.75\columnwidth]{GF.eps}
\includegraphics[width=1.0\columnwidth]{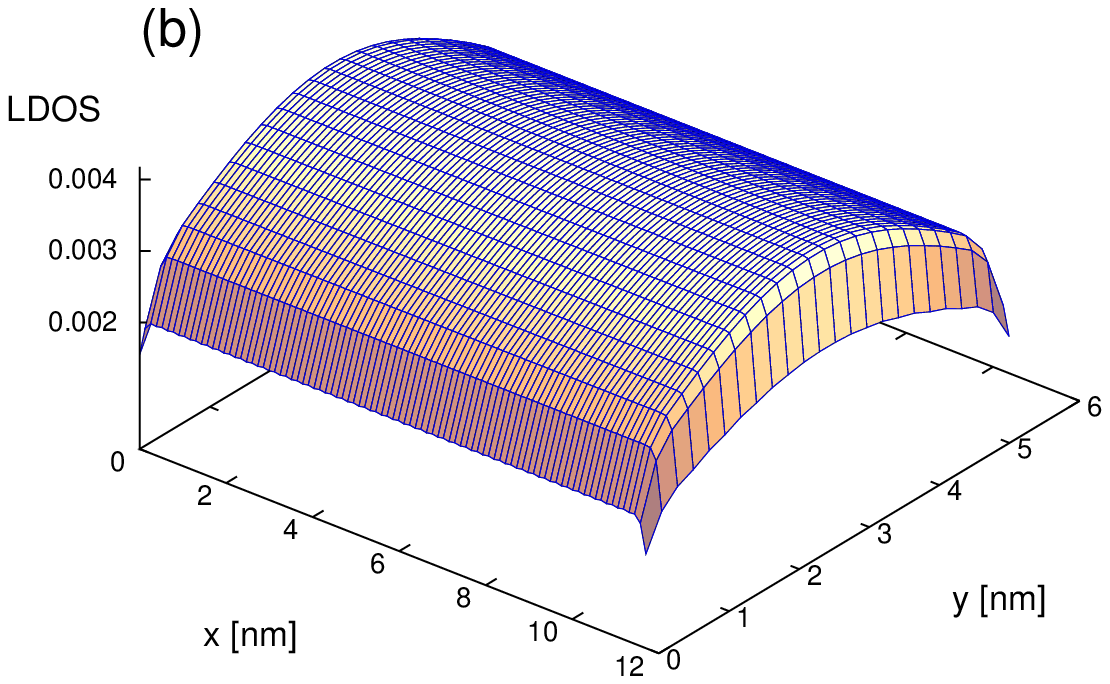}
\includegraphics[width=1.0\columnwidth]{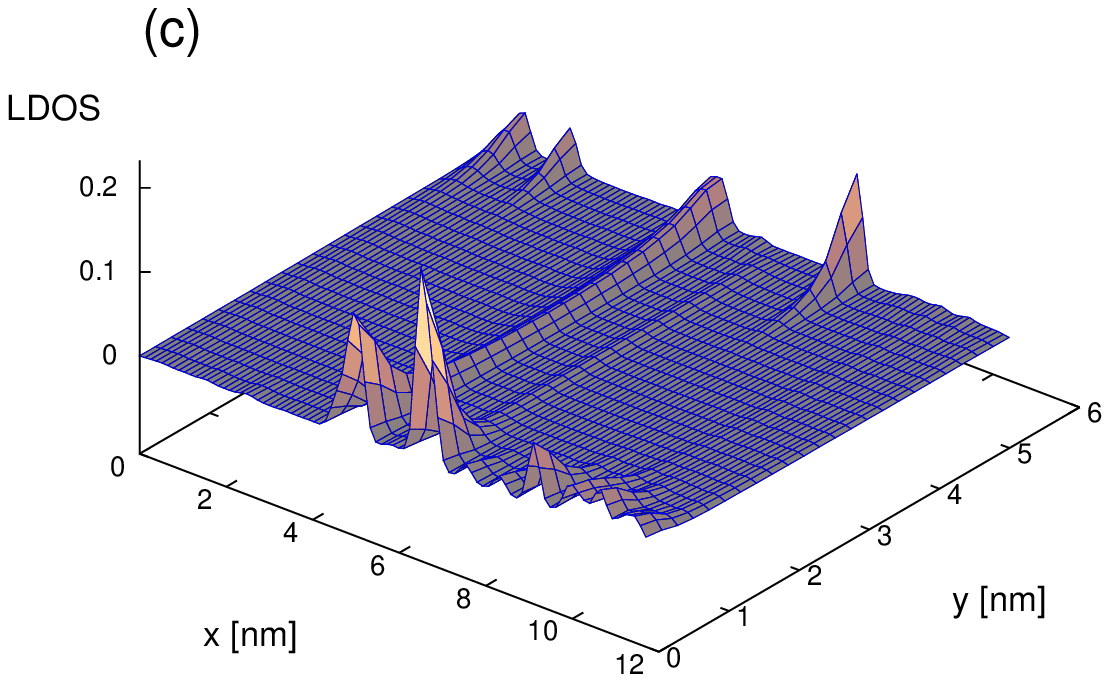}
\caption{(a) The linear conductance of a short nanoribbon (zigzag
  edges, $M=24$, $N=91$ sites) as a function of energy. The solid line
  corresponds to perfect edges while the dashed line corresponds to
  the edge disorder realization shown in the inset (three etching
  sweeps with $p_1=0.3$, $p_2=0.2$, $p_3=0.1$). Surface plot of the
  local density of states (arbitrary units, brickwall lattice
  representation) for the same nanoribbon at the energy value
  highlighted in (a): (b) clean case; (c) edge disordered case. Notice
  the appearance of localized states that traverse the nanoribbon when
  edge disorder is present. (The actual LDOS was convoluted with a
  Gaussian profile to smooth out high-frequency oscillations.)}
\label{fig:ldos_nanoribbon}
\end{figure}


The appearance of localized states in nanoribbons with edge disorder
is demonstrated in Fig. \ref{fig:ldos_nanoribbon} where the linear
conductance and the local density of states for a nanoribbon are shown
in the cases of perfect and irregular edges.

\begin{acknowledgements}
Financial support by the Brazilian funding agencies FAPERJ and CNPq is
gratefully acknowledged.
\end{acknowledgements}

\appendix

\section{Steps in the linear conductance}
\label{sec:appendixA}

Let us show that the surface Green's function in Eq. (\ref{eq:gLsq})
leads to the expect steps in the linear conductance. For this purpose,
let us begin by noticing that, in the case of a square lattice lead,
only propagating modes yield a finite level width: For
$|E-\varepsilon_\nu| < 2 t_x$,
\begin{equation}
\tilde{\Gamma}_\nu = -2\, \mbox{Im} \left[ \tilde{\Sigma}_\nu \right]
= 2\, \mbox{Im} \left[ \left( \tilde{g}_\nu \right)^{-1} \right] =
2\,t_x\, \sin \phi_\nu,
\end{equation}
where $\sin\phi_\nu = \sqrt{1 - (E-\varepsilon_\nu)^2/4t^2}$, in which
case we can write $\tilde{g}_\nu = e^{-i\phi_\nu}/t_x$.

In order to obtain the retarded Green's function across the system, we
add one slice between the left and right contacts and use the
following expression, easily derivable from Eqs. (\ref{eq:GLrec}),
(\ref{eq:GLlong}), (\ref{eq:Gexact}) and (\ref{eq:G0n}):
\begin{equation}
G_{0,2} = t_x^2\, g_L^2 \left( g_L^{-1} - t_x^2\, g_L \right)^{-1}
\end{equation}
Since $G_{0,2}$ depends solely on $g_L$, we can rewrite in the
propagation mode basis, in which case the Landauer formula is reduced
to [see Eq. (\ref{eq:caroli})]
\begin{equation}
{\cal T} = {\sum_{\nu}}^\prime \,\tilde{\Gamma}_\nu\, \left(
\tilde{G}_{0,2} \right)_\nu\, \tilde{\Gamma}_\nu\, \left(
\tilde{G}_{0,2} \right)_\nu^\ast,
\end{equation}
where the prime indicates that the sum runs only over states such that
$|E-\varepsilon_\nu| < 2 t_x$ and
\begin{equation}
\left( \tilde{G}_{0,2} \right)_\nu = t_x^2\, \tilde{g}_\nu^2 \left(
\tilde{g}_\nu^{-1} - t_x^2\, \tilde{g}_\nu \right)^{-1} = \frac{e^{-2i
    \phi_\nu}}{2i\, t_x\, \sin \phi_\nu}.
\end{equation}
Putting all together, we find that
\begin{align}
{\cal T} = & {\sum_{\nu}}^\prime \;\tilde{\Gamma}_\nu^2\, \left| \left(
\tilde{G}_{0,2} \right)_\nu \right|^2 \nonumber \\
= &{\sum_\nu}^\prime \; 1 = \#\ {\rm propagating\ modes\ for\ a\ given}\ E,
\end{align}
which is the expected result for a clean ballistic system.

\section{Peierls hopping phases}
\label{sec:A}

Here we evaluate the phase of the hopping matrix elements between any
two arbitrary sites due to the presence of a perpendicular magnetic
field. We pick the vector potential in the generic Landau gauge $A_x =
(\alpha-1) B y$ and $A_y = \alpha Bx$, with $0 \leq \alpha \leq
1$. The (directional) Peierls phase between two neighboring sites $k$
and $k^\prime$ is given by \cite{peierls}
\begin{eqnarray}
\varphi_{k,k^\prime} & = & \frac{ec}{\hbar} \int_k^{k^\prime} {\bf A}
\cdot d{\bf l} \nonumber \\ & = & \frac{ecB}{\hbar} \left[ (1-\alpha)
  \cos\theta_{kk^\prime} \int_{y_k}^{y_{k^\prime}} y\, dl + \alpha
  \sin \theta_{kk^\prime} \int_{x_k}^{x_{k^\prime}} x\, dl \right]
\nonumber \\ & = & \frac{ecB}{\hbar} \left[ (1-\alpha)\,
  \mbox{cotan}\, \theta_{kk^\prime} \int_{y_k}^{y_{k^\prime}}y\, dy +
  \alpha \tan \theta_{kk^\prime} \int_{x_k}^{x_{k^\prime}} x\, dx
  \right] \nonumber \\ & = & \frac{ecB}{\hbar} \left[ (1-\alpha)
  (x_{k^\prime} - x_k) \left( \frac{y_{k^\prime} + y_k}{2} \right)
  \right.\nonumber\\ && \hskip1.1cm \left.+ \alpha (y_{k^\prime} -
  y_k) \left( \frac{x_{k^\prime} + x_k}{2} \right) \right],
\end{eqnarray}
where $\theta_{kk^\prime}$ is the angle that the segment
$k$--$k^\prime$ makes with the $x$ axis. 

Notice that $\varphi_{k,k^\prime} = - \varphi_{k^\prime,k}$. If we sum
over all the bond phases around the perimeter of a hexagon, we obtain
$\sum \varphi = \sqrt{3}ecBa_0^2/2\hbar = 2\pi\, (\Phi/\Phi_0)$, where
$\Phi_0 = h/ec$ (flux quantum), and $\Phi = B\, A_{\rm hex}$, with
$A_{\rm hex} = \sqrt{3}\, a_0^2/2$ being the area of the hexagon.

\section{Random Flux Estimate}
\label{sec:flux}

Let us estimate the rms value of the random magnetic field produced by
the random vector potential:
\begin{eqnarray}
\left\langle B^2 \right\rangle & = & \left\langle (\partial_x A_y -
\partial_y A_x)^2 \right\rangle \nonumber \\ & = & \left\langle
\left\{ \frac{f}{\cal C} \sum_k \left[ c^x_{k}
\frac{2(x_{n,j}-X_k)}{\xi^2} + c^y_{k} \frac{2(y_{n,j}-Y_k)}{\xi^2}
\right]\, e^{-|{\bf r}_{n,j} - {\bf R}_k|^2/\xi^2} \right\}^2
\right\rangle \nonumber \\ & = & \frac{4f^2}{{\cal C}^2\xi^4} \sum_k
|{\bf r}_{n,j} - {\bf R}_k|^2\, e^{-2|{\bf r}_{n,j} - {\bf
R}_k|^2/\xi^2} \nonumber \\ & \approx & \frac{4f^2}{{\cal C}^2\xi^4
a_g^2} \int d^2R\, R^2\, e^{-2R^2/\xi^2} \nonumber \\ & \approx &
\frac{\pi\, f^2}{2\, {\cal C}^2a_g^2} = \frac{f^2 a_g^2}{2\pi\, \xi^4}.
\end{eqnarray}
%

%




\end{document}